\documentclass[
  aps,                 % American Physical Society style
  pra,                 % Physical Review A
  reprint,             % two-column, approximates the printed PRA layout
  superscriptaddress,  % numbered superscripts for the 4 different affiliations
  amsmath,amssymb,     % load AMS packages through the class (avoids clashes)
  floatfix,            % emergency float placement
  nofootinbib          % footnotes stay at the page bottom, not in the ref list
]{revtex4-2}

\usepackage{graphicx}
\usepackage{bm}
\usepackage{dcolumn}
\usepackage{hyperref}

\IfFileExists{orcidlink.sty}
  {\usepackage{orcidlink}}
  {\providecommand{\orcidlink}[1]{}}

\newcommand{\ano}{a_{e}}
\newcommand{\Anet}{\mathcal{A}}
\newcommand{\nhat}{\hat{\bm n}}
\newcommand{\aperp}{\bm{a}_{\perp}}

\newcommand{\figorbox}[2]{%
  \IfFileExists{#1}%
    {\includegraphics[width=\columnwidth]{#1}}%
    {\fbox{\begin{minipage}[c][0.62\columnwidth][c]{0.94\columnwidth}%
       \centering\footnotesize\ttfamily
       [artwork placeholder]\\[0.6ex]#1\\[0.6ex]\rmfamily\itshape #2
     \end{minipage}}}%
}

\begin{document}

\title{Net electron spin rotation in a plane-wave pulse:
       Holonomy set by the anomalous magnetic moment}

\author{N.~S. Akintsov\,\orcidlink{0000-0002-1040-1292}}
\email[Corresponding author: ]{akintsov777@ntu.edu.cn}
\affiliation{School of Artificial Intelligence and Computer Science,
             Nantong University, Nantong 226019, China}

\author{A.~P. Nevecheria\,\orcidlink{0000-0001-6736-4691}}
\email{artiom.nevecherya@gmail.com}
\affiliation{Department of Mathematical and Computer Methods,
             Kuban State University, Krasnodar 350040, Russia}

\author{S.~N. Andreev\,\orcidlink{0000-0003-3588-2894}}
\email{andreev@cir-innovations.ru}
\affiliation{Joint-Stock Company ``Center for Research and Development'',
             Moscow 101000, Russia}

\author{Qing-Hua Qin\,\orcidlink{0000-0003-0948-784X}}
\email{qinghua.qin@smbu.edu.cn}
\affiliation{Institute of Advanced Interdisciplinary Technology,
             Shenzhen MSU-BIT University, Shenzhen 518172, China}

\date{\today}

\begin{abstract}
We compute the spin rotation that survives after a relativistic electron has
crossed a plane-wave laser pulse of finite duration. In the interaction
picture built on the exact $g=2$ evolution, the
Thomas--Bargmann--Michel--Telegdi equation becomes parallel transport by a
connection with constant coefficients on the polarization plane, and the pulse
enters only through the closed curve that the transverse vector potential
traces there. The net rotation is the holonomy of that connection: an angle
$-\tfrac{1}{2}a_{e}^{2}\mathcal{A}$ about the propagation direction, with
$a_{e}=(g-2)/2$ the anomaly and $\mathcal{A}$ twice the signed area enclosed
by the curve. That area is the spin angular momentum the pulse carries per
unit area, so the rotation measures the helicity of the light. Reduction of
the residual dynamics to a rotation coupled through the anomaly alone is an
exact result of the 1960s [Ternov, Bagrov, and Klimenko, Sov.\ Phys.\ J.\ 11,
29 (1968); Bagrov and Gitman, \textit{The Dirac Equation and its Solutions}
(De Gruyter, Berlin, 2014), Sec.~5.3], which yields two closed-form
cases; the area law is the general second order that those two cases bound.
The same area governs the orientation memory of a neutral magnetic dipole
[Oblak and Seraj, Phys.\ Rev.\ D 109, 044037 (2024)] with a coupling
of order unity. For a charged electron on a Volkov orbit the coupling $g/2$
cancels identically, which suppresses the rotation by $1.3\times10^{-6}$ and
leaves a channel with no $g$-independent part. We verify the cancellation at
$g=2$ over 180 pulse configurations and the area law over 89 more. Finite
focusing restores that part at second order in $1/kw_{0}$, and it exceeds the
anomalous signal unless $w_{0}\gtrsim16\lambda$ at $\gamma=10$, or
$270\lambda$ at $\gamma=1$.
\end{abstract}

\maketitle

\section{Introduction}
\label{sec:intro}

Spin polarization of electrons in intense laser fields is usually treated as a
radiative process: the spin flips when a photon is emitted, and the flip rate
follows the structure of the
field~\cite{DelSorbo2017,Seipt2018,Li2019,Ivanov2004,Gonoskov2022,Fedotov2023}.
That channel closes as the quantum parameter $\chi$ becomes small, its
probability per pulse falling as $\alpha a_{0}\chi^{2}N$ for $N$ cycles at
normalized amplitude $a_{0}$. The spin keeps moving below it. Inside a pulse
with $a_{0}\sim10$ the Thomas--Bargmann--Michel--Telegdi (BMT)
equation~\cite{Bargmann1959} turns the rest-frame polarization through
radians, and nearly all of that rotation is undone while the pulse passes.

What survives is a holonomy. For a plane-wave background,
$A^{\mu}=A^{\mu}(\eta)$ with $\eta=k\cdot x$ the light-front phase, and for a
pulse of finite duration,
the rotation surviving after the field has gone is
$\Theta_{\rm net}=-\tfrac{1}{2}\ano^{2}\Anet$, where $\ano=(g-2)/2$ is the
anomalous magnetic moment and $\Anet$ is twice the signed area that the vector
potential encloses in the polarization plane. Every property of the pulse acts
through the curve that $\aperp(\eta)$ traces in that plane, and at leading
order through its area alone. The electron energy, the carrier frequency, the
carrier-envelope phase and the rate at which the curve is traced drop out.
Chirp and envelope shape do not: they deform the curve, and $\Anet$ follows.

The area carries a physical name. In the Coulomb gauge the spin angular
momentum density of the field is $\varepsilon_{0}\bm{E}\times\bm{A}$, and for a
plane wave $\bm{E}\propto-\aperp'$ while $\bm{A}\propto\aperp$, so that
$\Anet$ is the spin angular momentum the pulse carries per unit area. The net
spin rotation of the electron therefore measures the helicity of the light,
weighted by the square of the anomaly. Section~\ref{sec:holonomy} makes that
statement quantitative.

Two of the three ingredients of this result are in print, in two literatures
that have never cited one another, and Sec.~\ref{sec:prior} sets out the
division in full. The exact spin problem for an electron with an anomalous
moment in a plane wave was solved by the Tomsk school in the
1960s~\cite{Ternov1968,Chakrabarti1968,Bagrov1990,Bagrov2014} and rediscovered
in a gravitational context three decades later~\cite{Balakin2002}: the residual
dynamics is a rotation whose coupling is the anomaly alone, the first order in
the anomaly is algebraic in $\aperp(\eta)$, and exactly two profiles integrate
in closed form. Separately, the net reorientation of a precessing magnetic
dipole after a burst of radiation has passed is known to be of second order in
the coupling and to equal the helicity of the burst~\cite{Oblak2024}, a
statement phrased as a holonomy in the memory
literature~\cite{SerajNeogi2023}. Neither line supplies the general second
order for a pulse of arbitrary shape, and neither treats a charged relativistic
particle, where the reduction of the coupling from $g/2$ to $\ano$ is the whole
of the physics. Substituting an electron into the dipole formula overestimates
the answer by six orders of magnitude.

The rotation axis is the propagation direction in the electron rest frame, so
in a head-on or copropagating geometry the helicity is conserved through
$O(\ano^{6})$ and the measurable quantity is the rotation of a transverse
polarization relative to the momentum. At oblique incidence that axis leaves
the momentum and the helicity statement lapses. Linear polarization gives no
net rotation at any $g$, exactly and at every order, because the curve
degenerates to a segment.

Sections~\ref{sec:transport} and~\ref{sec:holonomy} carry the derivation.
Section~\ref{sec:prior} is the delimitation against earlier work.
Section~\ref{sec:numerics} reports the numerical verification,
Sec.~\ref{sec:focusing} the departure from the law in a focused beam, and
Sec.~\ref{sec:budget} the signal-to-background budget, which is unfavorable and
is given in full. Technical material, the validation tables and the extended
algebra are in the Supplemental Material, reproduced in full in the
Appendix of this preprint.

\section{Spin transport in a plane-wave pulse}
\label{sec:transport}

\subsection{Background, orbit and assumptions}

The background is an exact plane wave in the metric $(+,-,-,-)$: the
dimensionless potential $a^{\mu}(\eta)=eA^{\mu}(\eta)/m$ depends on the
light-front phase alone, with $k^{2}=0$ and $k\cdot a=0$. Its transverse part
is $\aperp$, with components $(a_{1},a_{2})=(a_{x},a_{y})$ along a
right-handed polarization basis $\hat{e}_{1}\times\hat{e}_{2}=\nhat$. The
amplitude parameter is $a_{0}$. It is the peak of $|\aperp|$ for linear
polarization, while for the elliptical family of ellipticity $\delta$ scanned
below that peak is $a_{0}/\sqrt{1+\delta^{2}}$, or $a_{0}/\sqrt{2}$ in the
circular case. Finiteness of the pulse enters through one condition,
\begin{equation}
  \aperp(+\infty)=\aperp(-\infty)=0 ,
  \label{eq:asympt}
\end{equation}
which is the statement that the transverse electric field has no dc component.
The phase average of $\aperp$ itself is unconstrained. A pulse carrying
$\int\aperp\,d\eta\neq0$ obeys everything below.

We treat the spin as a classical four-vector $S^{\mu}$ with $S\cdot u=0$ and
$S^{2}=-1$, evolving by the BMT
equation~\cite{Bargmann1959,Thomas1926,Jackson1999},
\begin{equation}
  \frac{dS^{\mu}}{d\tau}
  = \frac{e}{m}\left[\frac{g}{2}F^{\mu\nu}S_{\nu}
    + \left(\frac{g}{2}-1\right)u^{\mu}
      \bigl(S_{\lambda}F^{\lambda\nu}u_{\nu}\bigr)\right] ,
  \label{eq:bmt}
\end{equation}
with no Stern--Gerlach force, no radiation reaction and no electric dipole
moment. The first omission is the mildest: relative to the Lorentz force the
Stern--Gerlach force carries a factor $\hbar\omega/mc^{2}\simeq3\times10^{-6}$
for Ti:sapphire light in the rest frame of a slow electron, and its role in
classical electron dynamics has been isolated
numerically~\cite{Wen2016,Wen2017}. The other two are genuine restrictions.
Section~\ref{sec:budget} weighs them.

Within the plane-wave sector Eq.~(\ref{eq:bmt}) is not an approximation.
Chakrabarti solved the Dirac--Pauli equation exactly for fields of the form
$F=F(k\cdot x)$ with $k^{2}=0$ and found that the polarization four-vector obeys
the BMT equation exactly, for an arbitrary anomalous magnetic moment and an
arbitrary electric dipole moment~\cite{Chakrabarti1968}. Barducci and Giachetti
later built the eigenfunctions and the Green function of the same
problem~\cite{Barducci2008}. Outside the locally constant field approximation
the electron--wave interaction is nonlocal and an anomalous magnetic moment
cannot be introduced in the usual way~\cite{DiPiazzaPatuleanu2021}. We hold
$\ano$ at its vacuum value throughout. That costs a relative $O(\chi)$. It is
harmless at the working points of Sec.~\ref{sec:budget}.

The orbit is the exact Volkov solution~\cite{Volkov1935}. Because
$k_{\mu}F^{\mu\nu}=0$, the light-front projection $\kappa=k\cdot u$ is a
constant of the motion, $d/d\tau=\kappa\,d/d\eta$, and the four-velocity
returns to its initial value, $u(\pm\infty)=u_{0}$; no integration is needed
for the orbit, which removes one source of numerical error from everything
below.

\subsection{A single antisymmetric generator}
\label{sec:generator}

Write $(p\wedge q)^{\mu}{}_{\nu}=p^{\mu}q_{\nu}-q^{\mu}p_{\nu}$ for the wedge
of two four-vectors, which acts on $S$ as $p\,(q\cdot S)-q\,(p\cdot S)$; the
symbol never denotes a three-dimensional wedge. Every sign statement below is
read off that convention. With $f=(e/m)F=k\wedge a'$ and
$w^{\mu}=f^{\mu\nu}u_{\nu}$,
and using $u\cdot S=0$ to bring the right-hand side of Eq.~(\ref{eq:bmt}) to
antisymmetric form, one obtains a single generator,
\begin{equation}
  \frac{dS^{\mu}}{d\eta}=\frac{1}{\kappa}\,\Omega^{\mu}{}_{\nu}S^{\nu},
  \qquad
  \Omega=(1+\ano)\,k\wedge a' + \ano\,(u\wedge w) .
  \label{eq:generator}
\end{equation}
The completion to antisymmetric form is unique, because two antisymmetric
matrices that act identically on the three-space orthogonal to $u$ differ by an
antisymmetric matrix with a three-dimensional kernel, and the rank of an
antisymmetric matrix is even.

Expanded in the basis $\{k,\bar{k},e_{1},e_{2}\}$ with $k\cdot\bar{k}=2$, the
anomalous piece $u\wedge w$ has components along the boost $\bar{k}\wedge k$
and along $\bar{k}\wedge e_{i}$ and also inside the little algebra of $k$.
Hence at $g\neq2$ the generator leaves that little algebra. The pieces together
generate $\mathfrak{sl}(2,\mathbb{R})$ already for linear polarization, and the
$\eta$ ordering in Eq.~(\ref{eq:generator}) is not degenerate.

One simplification comes first. A null rotation $N$ from the little group
$ISO(2)_{k}$ leaves $k$ fixed and sends $e_{i}\mapsto e_{i}+c_{i}k$, so that
$NfN^{-1}=k\wedge(a'+ck)=f$: the background is invariant, which is stronger
than gauge equivalence. Since $ISO(2)_{k}$ acts transitively on the mass shell
at fixed $\kappa$, any $u_{0}$ can be carried to $(k+\bar{k})/2$, the net map
of the original geometry is the conjugate $N^{-1}\Lambda_{\rm net}N$, and a
rotation angle is a conjugation invariant. It is therefore enough to work
head-on, normalized to $\kappa=1$, and Secs.~\ref{sec:holonomy}
and~\ref{sec:numerics} do so.

\section{Reduction to a connection, and the holonomy}
\label{sec:holonomy}

\subsection{The $g=2$ propagator}
\label{sec:g2}

At $\ano=0$ the generator is $\Omega=k\wedge a'=a_{1}'N_{1}+a_{2}'N_{2}$ with
$N_{i}=k\wedge e_{i}$. Direct multiplication gives
$(N_{i}N_{j})^{\mu}{}_{\nu}=\delta_{ij}k^{\mu}k_{\nu}$, which is symmetric in
$(ij)$, so $[N_{1},N_{2}]=0$ and $N_{i}^{3}=0$; the generator therefore
commutes with itself at different phases, the ordering collapses, and
\begin{equation}
  \Lambda_{0}(\eta)=\exp\bigl[\kappa^{-1}(a_{1}N_{1}+a_{2}N_{2})\bigr],
  \qquad \Lambda_{0}(+\infty)=\mathbb{1} ,
  \label{eq:lambda0}
\end{equation}
by Eq.~(\ref{eq:asympt}). Hence no property of $u_{0}$ enters: $u$ does not
appear in $\Omega$ at $\ano=0$ at all. The mechanism is not ours. It is the one
Kupersztych identified: the operator that transports an electron and its spin
through a
plane wave is of Lorentz type and acts on the wave as a gauge
transformation~\cite{Kupersztych1976}. In modern language the generator sits in
the Abelian translation subgroup of the little group of a null
vector~\cite{Wigner1939}, and Eq.~(\ref{eq:asympt}) closes the loop.

\subsection{Interaction picture and the exact reduction}
\label{sec:reduction}

Since $(1+\ano)f+\ano\,u\wedge w=f+\ano\,(f+u\wedge w)$, factoring the evolution
as $U=\Lambda_{0}V$ gives an interaction-picture equation in which the anomaly
is the only coupling,
\begin{equation}
  \frac{dV}{d\eta}=\ano\,M(\eta)\,V ,
  \qquad
  M={\rm Ad}_{\Lambda_{0}^{-1}}\bigl(f+u\wedge w\bigr) .
  \label{eq:interaction}
\end{equation}
Write $A=k\wedge\aperp=a_{1}N_{1}+a_{2}N_{2}$, so that $\Lambda_{0}=e^{A}$ and
$M=e^{-{\rm ad}_{A}}X$ with $X=f+u\wedge w$. The series terminates after three
terms: on the whole of $\mathfrak{so}(1,3)$ the operator $A$ raises the grading
$\bar{k}\wedge e_{i}\to\{\bar{k}\wedge k,\,e_{1}\wedge e_{2}\}
\to k\wedge e_{i}\to0$,
so ${\rm ad}_{A}^{3}=0$ and $M=X-[A,X]+\tfrac12[A,[A,X]]$.

Two cancellations follow, and they are of different kinds. Everything along the
boost $\bar{k}\wedge k$ and along $e_{1}\wedge e_{2}$ is removed by the first
commutator, with no input beyond the algebra, while the residual null rotations
along $k\wedge e_{i}$ are removed by the two algebraic identities
\begin{equation}
  \mathcal{U}a_{1}-\mathcal{W}a_{2}=|\aperp|^{2}a_{1}' ,
  \qquad
  \mathcal{U}a_{2}+\mathcal{W}a_{1}=|\aperp|^{2}a_{2}' ,
  \label{eq:identities}
\end{equation}
with $\mathcal{U}=\aperp\!\cdot\!\aperp'$ and
$\mathcal{W}=a_{1}a_{2}'-a_{2}a_{1}'$, which follow from those definitions by
inspection. What survives is a pure rotation of the rest frame about an axis
lying in the polarization plane,
\begin{equation}
  M(\eta)\,d\eta=\nhat\wedge d\aperp,
  \qquad
  \frac{d\bm{\zeta}}{d\eta}
  = \ano\,\bigl(\nhat\times\aperp'\bigr)\times\bm{\zeta} ,
  \label{eq:connection}
\end{equation}
with $\bm{\zeta}$ the rest-frame spin and
$\nhat=\tfrac{1}{2}(k/\kappa-\kappa\bar{k})$ the unit spacelike vector along
the wave in the rest frame of $u_{0}$. In the convention fixed at
Eq.~(\ref{eq:generator}) the sign in Eq.~(\ref{eq:connection}) is positive, the
curvature is $[\nhat\wedge e_{1},\nhat\wedge e_{2}]=+\,e_{1}\wedge e_{2}$, and
$e_{1}\wedge e_{2}$ generates a right-handed rotation about $\nhat$. Table~S1
of the Supplemental Material carries the three terms of
$e^{-{\rm ad}_{A}}X$ out at one numerical point and shows where each
cancellation happens, and we also checked Eq.~(\ref{eq:connection}) against the
components of Eq.~(\ref{eq:generator}) on a phase grid, where the largest
discrepancy is $2.7\times10^{-15}$.

Equation~(\ref{eq:connection}) is a connection one-form on the polarization
plane, and its coefficients are constants. Two consequences are immediate. The
phase enters only through the path, so the ordered exponential of $\ano M$
depends on the curve $\mathcal{C}:\eta\mapsto(a_{x},a_{y})$ alone and not on
the rate at which it is traced. And the integral of a one-form with constant
coefficients along a path depends on the endpoints alone, so
\begin{equation}
  \ano\!\int\! M\,d\eta=\ano\,\nhat\wedge\Delta\aperp=0
  \label{eq:firstorder}
\end{equation}
by Eq.~(\ref{eq:asympt}), exactly, at any $a_{0}$ and for any curve. Hence the
first order in the anomaly cancels for every polarization. A weaker fact is
sometimes offered in its place and does not do the work. That
$\Lambda_{\rm net}$ lies in the stabilizer $SO(3)_{u_{0}}$ follows from
$u(\pm\infty)=u_{0}$ and the conservation of $S\cdot u$, but it confines the
first-order coefficient to a three-dimensional space and constrains it no
further.
Section~\ref{sec:focusing} makes the distinction concrete.

For linear polarization the axis $\nhat\times\aperp'$ is fixed, the rotations
at different phases commute, and
$V(+\infty)=\exp[\ano\,\Delta a_{1}\,\nhat\wedge e_{1}]=\mathbb{1}$ exactly at
any $g$ and any $a_{0}$. The conjugation of Sec.~\ref{sec:generator} carries
that conclusion to any geometry, since $N^{-1}\mathbb{1}N=\mathbb{1}$. This is
perturbative in nothing, and it agrees with the closed-form spin dynamics
obtained for intense linearly polarized fields~\cite{Walser2002}.

\subsection{Non-Abelian Stokes evaluation}
\label{sec:stokes}

The base is the polarization plane $\mathbb{R}^{2}$, the bundle is the trivial
principal bundle $\mathbb{R}^{2}\times SO(3)_{u_{0}}$, and $M$ lies in the
corresponding Lie algebra because $Mu_{0}=0$. The connection one-form is
$\omega=\ano(G_{1}\,da_{1}+G_{2}\,da_{2})$ with $G_{i}=\nhat\wedge e_{i}$, and
its curvature is constant over the plane,
\begin{equation}
  F=d\omega+\omega\wedge\omega
   =\ano^{2}\,(e_{1}\wedge e_{2})\,da_{1}\wedge da_{2} .
  \label{eq:curvature}
\end{equation}
The path $\mathcal{C}$ closes for exactly one reason,
Eq.~(\ref{eq:asympt}), and that is the only place where finiteness of the pulse
is used. Thus the net map is the holonomy
$\Lambda_{\rm net}=\mathcal{P}\exp\oint_{\mathcal{C}}\omega$. No adiabatic
parameter appears and no eigenspace is being followed, so this is ordinary
non-Abelian parallel transport, and the resemblance to the geometric phases of
Refs.~\cite{Berry1984,Wilczek1984} is an analogy.

The surface-ordered form of the non-Abelian Stokes
theorem~\cite{Halpern1979,Broda2001} turns the ordered exponential around
$\mathcal{C}$ into a surface-ordered exponential of the curvature carried back
to a base point along a mesh of paths. Constancy of $F$ in the coordinates of
the base does not make that ordering trivial, because what is ordered is not
$F$ but its adjoint transport along the mesh. Hence the theorem delivers the
leading term and nothing more: a rotation about $\nhat$ through $\ano^{2}$
times the oriented area enclosed by $\mathcal{C}$. The second Magnus
term~\cite{Magnus1954,Blanes2009} gives the same result in three lines and
fixes the sign, and the two together give
\begin{equation}
  \Theta_{\rm net}=-\tfrac{1}{2}\ano^{2}\Anet
    \bigl[1+O(\ano^{2}a_{0}^{2})\bigr],
  \quad
  \Anet=\int\bigl(a_{x}a_{y}'-a_{y}a_{x}'\bigr)\,d\eta ,
  \label{eq:holonomy}
\end{equation}
in the right-handed triad $\hat{e}_{1}\times\hat{e}_{2}=\nhat$, and Sec.~S2 of
the Supplemental Material carries both evaluations out in
full.

Two definitions have to be pinned down before Eq.~(\ref{eq:holonomy}) can be
read literally. First, $\Anet$ is twice the signed area enclosed by
$\mathcal{C}$, counted with multiplicity, so a curve that winds $N$ times about
the origin contributes $N$ times over. Second, we take $\Theta_{\rm net}$ to be
the angle lifted continuously in $\ano$ from $\Theta_{\rm net}=0$ at $\ano=0$.
The angle read off an $SO(3)$ matrix is its reduction modulo $2\pi$, and the
two coincide while $|\Theta_{\rm net}|<\pi$.

The area is gauge invariant. Within the plane-wave sector the freedom left is
$a^{\mu}\to a^{\mu}+c^{\mu}$ with $c$ constant and $k\cdot c=0$, the gauge
function being $c\cdot x$. The field tensor $f=k\wedge a'$ does not see it.
The transverse part of $c$ translates $\mathcal{C}$ bodily in the polarization
plane, and the signed area of a closed curve does not change under a
translation. Equation~(\ref{eq:asympt}) removes the freedom altogether in any
case, since it picks out the one representative that vanishes at both ends of
the pulse.

The electric pulse area of the unipolar-pulse
literature~\cite{Aleksandrov2020,Rosanov2024} is a different functional:
$\int E\,dt$ is linear in the field and vanishes for every pulse obeying
Eq.~(\ref{eq:asympt}), while $\Anet$ is quadratic and stays finite there. For
linear polarization $\mathcal{C}$ degenerates to a segment, a segment encloses
no area, and $\Anet=0$. The vanishing there is exact at every order, by the
fixed-axis argument of Sec.~\ref{sec:reduction} and not by
Eq.~(\ref{eq:holonomy}). The same holds for any curve traced out and back, such
as $a_{y}\propto a_{x}^{2}$.

\subsection{The area as the spin angular momentum of the pulse}
\label{sec:spinAM}

In the Coulomb gauge the spin part of the electromagnetic angular momentum has
density $\varepsilon_{0}\bm{E}\times\bm{A}$. For a plane wave written in terms
of the dimensionless potential, $\bm{A}=(m/e)\aperp$ and
$\bm{E}=-\partial_{t}\bm{A}=-\omega(m/e)\aperp'$, so that
\begin{equation}
  \varepsilon_{0}\!\int\!\bigl(\bm{E}\times\bm{A}\bigr)\,dt
  =\varepsilon_{0}\Bigl(\frac{m}{e}\Bigr)^{\!2}\Anet\,\nhat .
  \label{eq:spinAM}
\end{equation}
Thus $\Anet$ is, up to a positive constant, the spin angular momentum the pulse
carries per unit cross-sectional area, and Eq.~(\ref{eq:holonomy}) reads: the
net rotation of the electron spin is the helicity of the pulse times the square
of the anomaly. The identification is not ours. It is the reading given to the
same functional in the orientation memory of a magnetic
dipole~\cite{Oblak2024}, where the area is the difference between the numbers
of right- and left-handed photons in the burst. Equation~(\ref{eq:spinAM})
also separates the present effect from the Pauli coupling
$\bm{\sigma}\cdot(\bm{E}\times\bm{A})$, which rotates the spin already at $g=2$
in two counterpropagating elliptically polarized beams~\cite{Bauke2014b} and is
quantum in origin.

\subsection{Order of the remainder, and a resummation}
\label{sec:remainder}

The order of the remainder in Eq.~(\ref{eq:holonomy}) is a theorem, and the
proof takes two lines. Grade $\mathfrak{so}(3)_{u_{0}}$ by calling $G_{1}$ and
$G_{2}$ odd and $e_{1}\wedge e_{2}$ even. The bracket respects that
$\mathbb{Z}_{2}$ grading, since $[{\rm odd},{\rm odd}]={\rm even}$ and
$[{\rm even},{\rm odd}]={\rm odd}$. Every term of the Magnus series is an
integral of $n$-fold nested brackets of $M$, and $M$ is odd, so the $n$th term
has parity $n$. Three consequences follow: the component of the exponent along
$\nhat$ carries only even powers of $\ano$, which is the $O(\ano^{4})$
correction to Eq.~(\ref{eq:holonomy}); the leading correction to the axis is
the third-order in-plane term, so that at fixed curve the axis is tilted out of
$\nhat$ by $O(\ano)$; and the change in the projection of the spin on $\nhat$
is quadratic in that tilt, hence $O(\ano^{6})$. The helicity statement of
Sec.~\ref{sec:intro} rests on the last of the three. That one is measured.
With the spin started along $\nhat$ and the anomaly inflated so that
$\ano a_{0}$ runs from $0.03$ to $0.5$, the change in the longitudinal
projection follows
$\Delta S_{\parallel}=-1.80\times10^{-3}(\ano a_{0})^{6}$, with local log-log
exponents from $6.00$ to $5.98$ at the small-anomaly end. The coefficient
repeats to $0.5\%$ at $a_{0}=1$, $3$ and $10$. The dependence is on the
product $\ano a_{0}$ alone. Over the same range the angle scales as
$\ano^{2.000}$ and the tilt $\psi$ of the axis out of $\nhat$ as
$\ano^{0.996}$, and the identity
$\Delta S_{\parallel}=-(1-\cos\Theta_{\rm net})\sin^{2}\psi$ holds to
$5\times10^{-3}$, so the sixth power is the product of the two factors the
grading predicts. Two limits attach to it. For a spin that does not start
along $\nhat$ the projection changes at $O(\ano^{3})$ instead, measured
exponent $3.00$, so helicity is conserved to sixth order as a property of the
helicity state. The coefficient itself was measured at $N=1$, $\gamma=1$ and
circular polarization. It does not carry to the working point $a_{0}=75$,
$N=32$ of Sec.~\ref{sec:budget}.

One curve does better than the order counting. On a plateau where $\mathcal{C}$
is a circle of radius $a_{0}$ traced at unit rate, $\nhat\times\aperp'$ has
fixed magnitude $a_{0}$ and rotates about $\nhat$ at unit rate, and
Eq.~(\ref{eq:connection}) becomes the magnetic-resonance problem for a rotating
field of strength $\ano a_{0}$ on exact resonance. In the corotating frame the
rotation vector is constant, of length $\sqrt{1+\ano^{2}a_{0}^{2}}$ and tilted
from $\nhat$ by $\arctan(\ano a_{0})$, and returning to the laboratory frame
subtracts one turn per unit phase,
\begin{equation}
  \Theta \simeq
  \Bigl(\sqrt{1+\ano^{2}a_{0}^{2}}-1\Bigr)\,\frac{\Anet}{a_{0}^{2}} .
  \label{eq:resum}
\end{equation}
Three conditions attach to Eq.~(\ref{eq:resum}): the idealized circle violates
Eq.~(\ref{eq:asympt}) and is therefore not an admissible pulse; composing two
rotations about different axes returns a single angle only for an integer
number of turns; and for a pulse with fronts the phase interval has to be read
as $\Anet/a_{0}^{2}$, which is the form written above, since reading it as the
plateau length instead costs $19\%$ at $a_{0}=1$, $\ano=0.05$. The residual
error is real. At $\ano a_{0}=0.3$ Eq.~(\ref{eq:resum}) is still off by
$0.1\%$, so it improves the asymptotics without becoming exact. The word exact
does not belong to it. The expansion
parameter is $\ano a_{0}$, not $\ano$. With the physical anomaly
$\ano a_{0}=1.16\times10^{-3}a_{0}$, and Eq.~(\ref{eq:holonomy}) stays accurate
to $10^{-2}$ up to $a_{0}\sim100$ for a circular curve. For a strongly
elliptical or chirped curve the remainder depends on the shape of
$\mathcal{C}$ and not on its area alone, and we have not bounded it there.

\section{Relation to earlier work}
\label{sec:prior}

Three separate lines of work reach parts of Eq.~(\ref{eq:holonomy}), and the
boundaries between them and the present paper are sharp enough to be drawn one
by one.

\subsection{The exact plane-wave spin problem}
\label{sec:priorTomsk}

The Dirac--Pauli equation for a particle with an anomalous moment in a plane
wave of arbitrary profile was solved exactly in the 1960s. Ternov, Bagrov and
Klimenko treated a circularly polarized wave and read off the shift of the spin
precession frequency caused by the anomaly~\cite{Ternov1968}. Chakrabarti
obtained the polarization directly in integrated form for a class of fields
$F=F(k\cdot x)$, with an anomalous magnetic moment and an anomalous electric
dipole moment both arbitrary~\cite{Chakrabarti1968}. The material is collected
in the monograph of Bagrov and Gitman, Sec.~5.3 of
Ref.~\cite{Bagrov2014} and Chap.~8 of the earlier edition~\cite{Bagrov1990},
and four of its statements bear directly on this work.

First, after the Volkov factor is separated the residual spinor equation is
driven by the anomalous moment alone, with the wave entering through
$\aperp'(\eta)$, which is the content of Eq.~(\ref{eq:interaction}) above, in
quantum form and three decades earlier. Second, the first order in the anomaly
is written there in closed form, and the operator is an algebraic function of
$\aperp(\eta)$ and not an integral of it, so that our
Eq.~(\ref{eq:firstorder}) is one line away from that expression, although the
asymptotic step and the statement about a pulse are not taken there. That step
is short, and it was not taken in that line. Third, linear polarization of
arbitrary profile is solved exactly, the evolution operator being a function of
the instantaneous $\aperp(\eta)$. For a pulse that operator is the identity,
which is the exact zero quoted in Sec.~\ref{sec:reduction}. Fourth, a potential
modulus with an arbitrary phase function is solved exactly, and the answer
depends on the total phase swept and not on the phase function, which is
Eq.~(\ref{eq:resum}) together with its independence of chirp. The same
reduction was found independently three decades later, for electromagnetic and
gravitational pp-wave backgrounds, by Balakin, Kurbanova and
Zimdahl~\cite{Balakin2002}. They state explicitly that nothing rotates at
$g=2$.

One thing that line does not contain is the general second order. Bagrov and
Gitman
say plainly that two cases are known for which the residual equation is solved
exactly, and those are the two just listed; neither an asymptotic limit for a
pulse of finite duration nor an observable net rotation is formulated anywhere
in that literature. The language throughout is one of wave functions, quantum
numbers and almost-periodicity. Equation~(\ref{eq:holonomy}) is the second
order for an elliptical curve of arbitrary profile, and the two exactly
solvable cases are its degenerate limits: for linear polarization the curve
collapses and $\Anet=0$, while for a circular plateau the area is known in
closed form and Eq.~(\ref{eq:holonomy}) is the leading term of
Eq.~(\ref{eq:resum}).

One consequence of that reading is worth recording. In the Dirac--Pauli problem
the anomalous magnetic and electric moments enter through a single modulus and
a constant mixing angle, and a constant rotation in the transverse plane leaves
the signed area unchanged. Equation~(\ref{eq:holonomy}) therefore holds for a
particle carrying both, with $\ano$ replaced by the modulus of the combined
anomalous moment.

\subsection{The vanishing at $g=2$}

That a plane wave leaves the asymptotic spin of a $g=2$ electron untouched is
old and well recorded. Kupersztych derived the underlying gauge mechanism in
1976~\cite{Kupersztych1976}. Aleksandrov \textit{et al.} found that the spin
change in a linearly polarized pulse is proportional to the electric pulse area
$\int E\,dt$~\cite{Aleksandrov2020}, which is zero for any pulse with a finite
envelope. For arbitrary polarization the same statement follows in one line
from the Volkov solution~\cite{Volkov1935}. Ilderton, King and Tang proved it in
the quantum theory and wrote the first order in the anomaly explicitly: the
amplitude for reversing the spin state is
$|c_{-}|=\bigl|\tfrac{1}{2}\int d\eta\,\mu_{b}\,(a_{x}'-ia_{y}')\bigr|$, with
$\mu_{b}$ their phase-dependent anomaly and the light-front integral running
over the whole pulse~\cite{Ilderton2020}. The result is stated verbatim in the
review literature~\cite{Gonoskov2022,Fedotov2023}. Section~\ref{sec:g2} adds
nothing to it and cites it as the starting point.

The delimitation against Ref.~\cite{Ilderton2020} needs care, because their
vanishing condition is parity: the amplitude is zero when the potential is an
even function of the phase. Unipolarity enters only in the weak-field limit.
For a pulse of zero electric area that is odd in the phase, produced by a
shifted carrier-envelope phase, a chirp or an asymmetric envelope, their first
order survives and can exceed the holonomy. By how much we have not evaluated.
A constant $\mu_{b}$ returns $\mu_{b}\Delta\aperp=0$ by
Eq.~(\ref{eq:asympt}), so what is left of their integral is the variation of
$\mu_{b}$ across the pulse, and that variation is itself of order $\chi$.
Their expression is first order in a dressed
anomaly $\mu_{b}(\chi)$. Ours is second order in the vacuum anomaly, and it
does not vanish when theirs does. The two are complementary. The frame common
to both, that non-radiative effects dominate at small $\chi$, is theirs.

\subsection{Area, helicity and memory}
\label{sec:priorMemory}

The functional $\Anet$ and its identification with helicity are also in print,
in a literature with no overlap with strong-field physics. Oblak and Seraj
computed the net reorientation of a magnetic dipole after a burst of radiation
has passed. They found a second-order term equal to the signed area swept by
the transverse potential, and they identify that area as the optical helicity
of the burst~\cite{Oblak2024}. Seraj and Neogi phrase memory observables of this
kind
as holonomies~\cite{SerajNeogi2023}. Up to the coupling constant, their
nonlinear term and Eq.~(\ref{eq:holonomy}) are the same expression, and we
take both the area law and the helicity reading from them, so that
Sec.~\ref{sec:spinAM} is written accordingly.

The physics is different, and the difference is the point. Their test body is a
neutral, static, nonrelativistic dipole obeying
$\dot{M}^{i}=\mathrm{k}\,\mathcal{F}^{ij}M_{j}$, and the coupling $\mathrm{k}$
is the full gyromagnetic ratio, a number of order unity: there is no charge, no
orbit, no Thomas precession and no $g$ factor. Substituting an electron into
that formula with $\mathrm{k}=g/2$ overestimates the rotation by a factor
$(g/2)^{2}/\ano^{2}=7.4\times10^{5}$. For a charged relativistic electron on an
exact Volkov orbit the terms proportional to $g/2$ cancel identically against
the boost and Thomas parts of the generator, as Eq.~(\ref{eq:interaction})
shows, and what remains couples through $\ano$ alone. Hence their result cannot
be applied to an electron before that reduction is proved, and ours does not
follow from theirs. The reverse implication also fails. Their formula carries a
term linear in the field, the radial magnetic field at null infinity, which is
identically zero for an exact plane wave.

The suppression is not a defect. A $g$-independent background of the same
geometric form is absent by Eq.~(\ref{eq:lambda0}), so in an ideal plane wave
the entire net rotation is anomalous. The factor $1.3\times10^{-6}$ is the
price of a channel with nothing to subtract. Section~\ref{sec:focusing} shows
how a finite waist puts the background back.

\subsection{Schemes that do rotate the spin}

Two published mechanisms produce a net rotation of the same observable and must
not be confused with this one. Wei \textit{et al.} rotate the transverse
polarization of a relativistic beam through an arbitrary angle with a
copropagating chirped or subcycle pulse~\cite{Wei2023}. Their pulses are
temporally asymmetric, which is $\aperp(+\infty)\neq\aperp(-\infty)$ and breaks
Eq.~(\ref{eq:asympt}) at order $\ano^{0}$. The dominant coupling in their
analytic estimate is the kinematic $1/\gamma$ and not the anomaly. Their result
is chirp dependent while ours is not, and there is no contradiction. What
depends on the chirp there is the accumulated precession phase, whereas the
invariant here is the net rotation relative to the momentum, that is, the
difference between the precession phase and the momentum phase. Tikhomirov
showed that a phase velocity $v_{\rm ph}\neq c$ produces a spin precession
exceeding radiative effects by orders of
magnitude~\cite{Tikhomirov2001,Tikhomirov2003}; that effect vanishes at
$v_{\rm ph}=c$, so it confirms the plane-wave theorem from the other side and
identifies a medium as a way of breaking it.

\subsection{What is new here}

Four items remain, and they are the claims of this paper. (i) The net rotation
for a pulse of finite duration and arbitrary elliptical profile is
$-\tfrac12\ano^{2}\Anet$, Eq.~(\ref{eq:holonomy}), which is the general second
order that the two known exact cases bound. (ii) The geometric reading:
holonomy of a connection with constant coefficients on the polarization plane,
a signed area evaluated by the non-Abelian Stokes theorem, and a
$\mathbb{Z}_{2}$ grading that fixes the remainder at $O(\ano^{4})$, the axis
tilt at $O(\ano)$ and the helicity change at $O(\ano^{6})$. (iii) The transfer
of the area mechanism from a neutral dipole to a charged relativistic electron,
where the correct coupling is the anomaly and not $g/2$. That is the
$1.3\times10^{-6}$ suppression, and it is also the absence of a $g$-independent
background.
(iv) Invariance under reparameterization of the phase, together with the
quantitative bound on finite focusing of Sec.~\ref{sec:focusing}. Everything
else in Secs.~\ref{sec:transport} and~\ref{sec:holonomy} we take from the works
named above.

\section{Numerical verification}
\label{sec:numerics}

\begin{figure}[tb]
  \figorbox{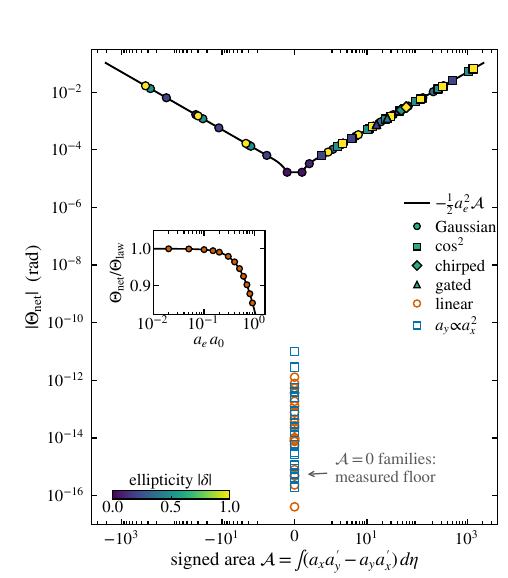}{FIG.~1 artwork: net rotation against signed area,
    one point per pulse configuration, plus the linear-polarization and
    zero-area families at the roundoff floor.}
  \caption{Net spin rotation in a plane-wave pulse against the signed area of
    the vector potential. Each symbol is one solution of Eq.~(\ref{eq:bmt})
    along the exact Volkov orbit, and the solid line is
    Eq.~(\ref{eq:holonomy}) with no fitted parameter. The scan covers Gaussian
    and $\cos^{2}$ envelopes, $N=0.5$--$16$ cycles, chirped and
    polarization-gated pulses, ellipticities $\delta=0.05$--$1$,
    carrier-envelope phases over $[0,2\pi)$ and Lorentz factors
    $\gamma=1$--$10^{4}$, with the anomaly set to $\ano=10^{-2}$ so that the
    rotation stands above the roundoff floor. The $g=2$ null family of the
    text is a separate set of runs covering
    $\gamma\le5\times10^{3}$. Color is the ellipticity $|\delta|$ of the
    pulse, and for the polarization-gated family that of the residual circular
    component, $|(1-q)/(1+q)|$ with $q$ the amplitude ratio of the two
    counter-rotating pulses; it adds nothing beyond the scan parameters. Open
    symbols are the two families
    for which $\Anet=0$: linearly polarized pulses, and the zero-area curve
    $a_{y}\propto a_{x}^{2}$ traced out and back. All $42$ points lie between
    $4.1\times10^{-17}$ and $9.95\times10^{-12}\,\mathrm{rad}$ for anomalies
    up to $\ano=0.6$, the two maxima being $1.29\times10^{-12}$ (linear) and
    $9.95\times10^{-12}$ (zero area).
    Inset: the ratio $\Theta_{\rm net}/\Theta_{\rm law}$,
    $\Theta_{\rm law}=-\tfrac{1}{2}\ano^{2}\Anet$, against $\ano a_{0}$ at
    $a_{0}=1$ for a circularly polarized flat-top pulse whose plateau is a
    circle of radius $a_{0}$, together with the resummation of
    Eq.~(\ref{eq:resum}).}
  \label{fig:holonomy}
\end{figure}

We integrated Eq.~(\ref{eq:bmt}) along the analytic Volkov orbit with the
DOP853 scheme at ${\rm rtol}=10^{-13}$ and ${\rm atol}=10^{-16}$, transporting
an orthonormal rest-frame triad and reading the net rotation off the resulting
matrix. Reading the angle from a transported frame is what makes the result
independent of the initial polarization, and a second implementation in
covariant bivector form, run at a looser tolerance, returns the same zeros on a
floor four to seven orders coarser. At the end of the pulse $|S\cdot u|$ and
$|S\cdot S+1|$ stay below $3\times10^{-12}$ at $\gamma=1$ in the first scan
and below $6\times10^{-11}$ in the set of Fig.~\ref{fig:holonomy}. Both floors
grow with $\gamma$, approaching $\gamma^{2}$ above $\gamma\simeq10^{2}$, while
the
measured $\Theta_{\rm net}$ at $g=2$ does not move with $\gamma$ at all.

Figure~\ref{fig:holonomy} collects the scan. Over 180 configurations at $g=2$,
covering two envelopes, $\gamma$ from $1$ to $5\times10^{3}$, $a_{0}$ from
$0.1$ to $10$, and three carrier-envelope phases, the largest net rotation is
below $3\times10^{-13}\,\mathrm{rad}$. At $a_{0}=10$ that is the roundoff
floor. Section~S4.3 of the Supplemental Material shows the
tolerance scan that establishes it. For linear polarization the same floor
persists once the anomaly is switched on and inflated by up to four orders of
magnitude above its physical value, and at $\gamma\le10^{2}$ the residue stays
at or below $4.1\times10^{-11}\,\mathrm{rad}$.

Away from linear polarization, Eq.~(\ref{eq:holonomy}) reproduces the measured
angle to three digits across the scan of Fig.~\ref{fig:holonomy} and to five in
the ellipticity scan. The ratio $\Theta_{\rm net}/\ano^{2}$ is constant to
$0.8\%$ across a $26$-fold range in $\ano$, the scaling
$\Theta_{\rm net}\propto\delta/(1+\delta^{2})$ in the ellipticity holds to five
digits, and the rotation grows linearly with pulse length. Envelope shape acts
through the area and through nothing else, and the effect is measurable: at
$a_{0}=10$ and $N=1$ the Gaussian gives $\Anet=334.412$ against $323.603$ for
$\cos^{2}$, a difference of $3.2\%$ that propagates straight into
$\Theta_{\rm net}$. Reparameterizing the phase at fixed curve leaves the net
rotation unchanged to eight or nine digits, even at the inflated value
$\ano=0.25$. That invariance holds by construction, so it tests the code and
not the theorem.

The absence of any $\gamma$ dependence is kinematic. A null rotation removes the
transverse initial momentum while leaving $k$, $\kappa$ and the field tensor
untouched, and $\kappa$ itself survives only as the normalization of $k$, that
is, as a reparameterization of the phase. Neither operation changes
$\aperp(\eta)$. Hence the only inputs left are the curve $\mathcal{C}$ and the
number $\ano$. Values at $\gamma=1$ and $\gamma=10^{4}$ agree to six digits.
The storage-ring enhancement $\ano\gamma$~\cite{Baier1972,Mane2005} compares
precession rates about a static field at one instant of laboratory time. Here
the quantity it would multiply is exactly zero.

At the physical anomaly the angles are small. A circularly polarized Gaussian
pulse gives
$\Theta_{\rm net}=2.25\times10^{-6}a_{0}^{2}N\,\mathrm{rad}$, against
$2.18\times10^{-6}a_{0}^{2}N\,\mathrm{rad}$ for the $\cos^{2}$ envelope.
Section~\ref{sec:budget} turns those numbers into a working point.

\section{Finite focusing}
\label{sec:focusing}

Equation~(\ref{eq:holonomy}) rests on $A=A(\eta)$. Every real beam has a waist,
and in a focused pulse the spin moves already at $g=2$. The failure of the
Lawson--Woodward theorem in a focused beam~\cite{Lawson1979,Esarey1995} leaves
a residual momentum transfer. The hierarchy of spin contributions in the
diffraction parameter $\varepsilon=1/(kw_{0})$ was set out qualitatively in
Ref.~\cite{Ilderton2020}, and a helicity change of up to $10\%$ during
ponderomotive scattering was reported in Ref.~\cite{Wen2022}. Neither work
gives the exponent.

We modeled a focused Gaussian pulse with the Lax--Louisell--McKnight expansion
carried to third order in $\varepsilon$, closing the Coulomb gauge at the
retained order~\cite{Lax1975,Salamin2002}, and integrated the Lorentz and BMT
equations in laboratory time. The Maxwell residuals of that model and the
systematic spread over its one free constant are tabulated in Sec.~S5 of the
Supplemental Material. The observable is the spin rotation
relative to the momentum: the exit reference triad is carried from the entrance
triad by the pure boost $u_{0}\to u_{f}$, so the kinematic transport that
accompanies ponderomotive deflection is subtracted. Five triads with anomalies
$\ano\in\{0,\pm h,\pm2h\}$, $h=0.05$, then share one orbit, and five-point
differences give the rotation vector as $r=r_{0}+\ano r_{1}+\ano^{2}r_{2}$.

\begin{figure}[tb]
  \figorbox{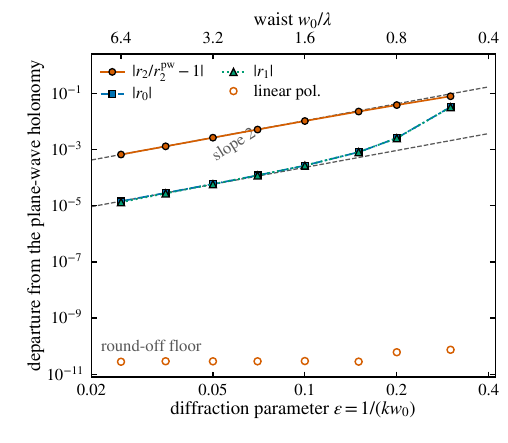}{FIG.~2 artwork: log-log plot of the three
    focusing-induced terms against the diffraction parameter, all with
    slope 2.}
  \caption{Departure from the holonomy for a focused Gaussian pulse against
    the diffraction parameter $\varepsilon=1/(kw_{0})$, for $a_{0}=1$, $N=1$,
    circular polarization, head-on collision at $\gamma=10$. Plotted are the
    relative deviation of the second-order coefficient $r_{2}$ from
    $-\tfrac{1}{2}\Anet\nhat$, and the magnitudes of the zeroth- and
    first-order coefficients $r_{0}$ and $r_{1}$, which vanish identically in
    a plane wave. All three follow $\varepsilon^{2}$ for
    $\varepsilon\lesssim0.15$. The dashed guide has
    slope 2, and the local log-log slopes measured over
    $\varepsilon=0.025$--$0.15$ run from $1.92$ to $2.00$ for the deviation of
    $r_{2}$ and from $2.71$ to $2.02$ for $r_{0}$. The upper axis gives the
    waist in wavelengths, $w_{0}/\lambda=1/(2\pi\varepsilon)$. Open circles
    repeat the deviation of $r_{2}$ for linear polarization, where it stays
    at or below the roundoff floor $7.5\times10^{-11}$ for every
    $\varepsilon$.}
  \label{fig:focusing}
\end{figure}

Every violation enters at order $\varepsilon^{2}$ for
$\varepsilon\lesssim0.15$, as Fig.~\ref{fig:focusing} shows. The second-order
coefficient keeps its plane-wave value with a relative error
$1.05\varepsilon^{2}$ at $\gamma=10$, rising to $3.4\varepsilon^{2}$ at
$\gamma=1$, and the local log-log slope of that error approaches $2.00$ as
$\varepsilon\to0$. Two terms that are identically zero in a plane wave grow
alongside it, $|r_{0}|$ and $|r_{1}|$, so both the $g$-independent rotation and
the first order in the anomaly are restored; their exponents fitted over the
whole plotted range are $2.87$ and $2.89$, with local slopes that fall to
$2.02$ at the small-$\varepsilon$ end, so that the quadratic law holds only as
an asymptote and not over the whole scan. A direct run at the physical anomaly
gives
\begin{equation}
  \frac{\Theta_{\rm anom}(\varepsilon)}{\tfrac{1}{2}\ano^{2}\Anet}-1
  \simeq \frac{10.7}{a_{0}}\,\varepsilon^{2}
  \qquad (\gamma=10) ,
  \label{eq:epsviolation}
\end{equation}
the coefficient being measured as $10.7$, $10.8$ and $13.5$ at $a_{0}=1$ and
$\varepsilon=0.025$, $0.050$ and $0.150$. Therefore Eq.~(\ref{eq:holonomy})
holds to $1\%$ for $w_{0}>5.2\lambda$ at $a_{0}\simeq1$ and $\gamma=10$. That
number carries the systematic uncertainty of the field model: a factor of three
on the coefficient, holding steady as $\varepsilon$ falls, so at the far end of
the spread the requirement becomes $w_{0}>9\lambda$. The exponent survives that
spread. At $\gamma=1$ the same accuracy needs $w_{0}>99\lambda$, and
$10\%$ accuracy still needs $w_{0}>31\lambda$; the waist requirement relaxes as
the intensity rises, as $1/a_{0}$ up to $a_{0}\simeq2$, and we have not tested
it above $a_{0}=24$.

However, the dominant focusing term carries no information about $g$. For
$\gamma\ge100$ we measure $|r_{0}|=|\Delta u_{\perp}|$ to four digits with the
axis orthogonal to the momentum, which identifies $r_{0}$ as the
Thomas--Wigner rotation~\cite{Thomas1926,Wigner1939} that accompanies the
ponderomotive kick~\cite{Mackenroth2019} and is fixed by the orbit alone. With
it go two properties of the plane-wave result: the net rotation is nonzero
already at $g=2$, and helicity is no longer conserved, because the rotation
axis has turned perpendicular to the momentum. At $\varepsilon=0.15$ and
$a_{0}=24$ the anomalous part amounts to a $5\times10^{-5}$ fraction of the
total signal. At $a_{0}=1$ the same ratio is $1.25\times10^{-3}$, twenty-five
times larger. One property does survive: for linear polarization $r_{2}$ stays
at the roundoff floor for every $\varepsilon$ up to $0.30$, although nothing in
Eq.~(\ref{eq:holonomy}) guarantees it there. We have no argument for why. All
of this is one electron on one orbit. We did not average over a beam, and the
deposited scans give $|r_{0}|$ varying by a factor of $20$ with impact
parameter inside a single waist.

\section{Signal, background and the classical window}
\label{sec:budget}

The estimates below use published scalings, and no radiative process was
integrated here, classically or quantum mechanically. The quantum nonlinearity
parameter is $\chi\simeq6.07\times10^{-6}\gamma a_{0}$ at
$\hbar\omega=1.55\,\mathrm{eV}$. The three limits on a classical treatment are
the radiated fraction $R_{C}\simeq\tfrac23\alpha a_{0}\chi N$, the photon
number $N_{\gamma}\sim\alpha a_{0}N$, and the radiative spin-flip probability
$P_{\rm sf}\sim\alpha a_{0}\chi^{2}N$~\cite{DiPiazza2012,Seipt2018,Li2019}.
All three tighten with $\gamma$ while $\Theta_{\rm net}$ does not move with it,
so the optimum runs against intuition: keep the electron slow and pay in
intensity and duration.

Take the best classical point of the plane-wave grid, $\gamma=1$, $a_{0}=75$
and $N=32$, an intensity of $1.2\times10^{22}\,\mathrm{W\,cm^{-2}}$ at
$800\,\mathrm{nm}$, where $\Theta_{\rm net}\simeq0.4\,\mathrm{rad}$ and the
transverse depolarization is $\sin^{2}(\Theta_{\rm net}/2)\simeq4\times10^{-2}$,
with $\chi\simeq2\times10^{-4}$, $R_{C}\simeq3\times10^{-3}$ and
$P_{\rm sf}\sim10^{-6}$. That probability sits four to five orders below the
depolarization it would have to compete with, and it acts differently: a
radiative flip removes one spin from the polarized fraction and leaves the
angle of the remainder untouched. Radiation reaction is the more dangerous of
the two, because it shifts the angle itself. Its weight can be estimated as a
relative distortion $2R_{C}$ of the curve traced in the rest frame against a
holonomy of relative size $\ano a_{0}$, giving
$\Theta_{\rm RR}/\Theta_{\rm net}\sim2R_{C}/(\ano a_{0})$. At the point just
quoted that estimator is $6\times10^{-2}$, and at $\gamma=10$, $a_{0}=75$,
$N=8$ it is $0.30$. A correction of six to thirty percent sits badly with a law
quoted to a percent, and only an integration of the Landau--Lifshitz equation
would settle the matter. We did not perform one. At $\gamma=10^{3}$ the same
estimator reaches $4$, and $P_{\rm sf}$ exceeds the geometric signal outright:
at $a_{0}=10$ and $N=8$ the two are $2\times10^{-3}$ and
$1.8\times10^{-3}\,\mathrm{rad}$. The correction has overtaken the quantity it
corrects.

The focusing background is the harder problem, and the numbers of
Sec.~\ref{sec:focusing} settle it against us. The criterion is
$|r_{0}|<\ano^{2}|r_{2}|$. Relative accuracy of $r_{2}$ is a different
question. At the softest
point of the $\varepsilon$ scan, $\varepsilon=0.025$ or $w_{0}=6.4\lambda$, at
$a_{0}=1$, $N=1$ and $\gamma=10$, the measured $|r_{0}|=1.44\times10^{-5}$
stands against an anomalous signal $\ano^{2}|r_{2}|=2.25\times10^{-6}$, so the
$g$-independent background exceeds the whole signal by a factor of $6.4$.
Hence pushing the background below the signal requires $\varepsilon^{2}<
\ano^{2}|r_{2}|/c_{0}$ with the measured $c_{0}$, that is $w_{0}>16\lambda$ at
$\gamma=10$ and $w_{0}>13.6\lambda$ at $\gamma=10^{3}$.

However, two axes save that comparison at high energy and destroy it at low.
For $\gamma\ge100$ the background axis is orthogonal to the momentum,
$|r_{0\parallel}|/|r_{0\perp}|\le6\times10^{-5}$ at $\gamma=10^{3}$, while the
holonomy lies along the momentum, so a measurement of the longitudinal
component separates the two. The separation is two orders of magnitude weaker
at $\gamma=10^{2}$, where the ratio is $5.7\times10^{-3}$, and it fails below
$\gamma\simeq3$: at $\gamma=1$,
$\varepsilon=0.15$ and $a_{0}=1$ the background is mostly parallel to the
momentum, $0.148$ against $0.014$ transverse, which is exactly the axis the
signal occupies. Requiring
$0.148\,(\varepsilon/0.15)^{2}<2.25\times10^{-6}$ then gives
$w_{0}>270\lambda$.

Therefore the two ends close on each other: at $\gamma\simeq1$ radiation is
harmless, but the focusing background shares the signal axis and exceeds it by
four to five orders of magnitude at any attainable waist, while at
$\gamma\gtrsim10^{2}$ the axes separate and the radiative flip and the radiated
fraction have overtaken the effect instead. No window is left between them. The
flagship point compounds the problem from the other side. Setting $a_{0}=75$ at
$w_{0}=31\lambda$, which is the $10\%$ row of the waist table at $\gamma=1$,
means a pulse power
$I\pi w_{0}^{2}/2\simeq1.2\times10^{17}\,\mathrm{W}$ and about $10\,\mathrm{kJ}$
in $32$ cycles, one to two orders above the largest systems in operation or
under construction. Focused calculations were run only at $N=1$ and
$a_{0}\le24$, so that point also lies outside the validated region. We do not
claim an experimental window, and we state the budget in full because the
alternative is to show the favorable half of each comparison.

Three further qualifications belong here. The obvious escape, a particle with a
larger anomaly, is closed by the mass: the relevant parameter is
$\ano a_{0}\propto\ano\,m_{e}/m$, which is $1.16\times10^{-3}a_{0}$ for the
electron against $9.8\times10^{-4}a_{0}$ for the proton and
$5.6\times10^{-6}a_{0}$ for the muon, so the proton anomaly of $1.793$ buys
nothing. Extracting $\ano^{2}$ from a measured angle is no better than the
absolute calibration of $a_{0}^{2}N$ inside the interaction volume, which at
$10^{22}\,\mathrm{W\,cm^{-2}}$ is a few tens of percent, and the focal average
was never taken. And the configuration needs a transversely polarized beam of
$\gamma\simeq1$--$10$ synchronized to the focus, for which laser-plasma sources
of polarized electrons are the natural
starting point~\cite{Buescher2020,Thomas2020,Reichwein2025}; whether the
required control of the impact parameter is realistic is an experimental
question we have not addressed.

\section{Scope and outlook}
\label{sec:outlook}

Every hypothesis behind Eq.~(\ref{eq:holonomy}) can be broken on purpose.
Photon emission and radiation reaction spoil $u(+\infty)=u_{0}$, and radiation
reaction brings in a dimensionful time, so $\gamma$ and $\omega$ reenter. The
closest published mechanism turns the spin toward the propagation axis of a
circularly polarized field at first order in $\ano$, and it is emission that
keeps the transverse spin phased there~\cite{Li2022}. Spin-dependent radiative
deflection acts in the same regime~\cite{Geng2020}. A standing wave adds a
second null direction, so that the two little algebras together generate the
whole Lorentz algebra and the ordering survives at
$g=2$~\cite{Bauke2014a,Rosanov2021}, while a plasma gives $k$ a mass,
$k^{2}=\omega_{p}^{2}$, and a first-order rotation returns with weight
$(n_{e}/n_{c})/(\ano a_{0})$, already of order $0.5$ at
$n_{e}/n_{c}=6\times10^{-3}$ and $a_{0}=10$. That rotation depends on the
electron energy and on the carrier frequency, where the holonomy does not, and
it disappears as $v_{\rm ph}\to c$~\cite{Tikhomirov2001,Tikhomirov2003}. In
addition, self-interaction in a strong plane wave carries a spin dynamics of
its own at the same order in $\alpha$~\cite{Meuren2011}. We integrated none of
these cases. The estimates come from the standard scalings collected in
Refs.~\cite{DiPiazza2012,Gonoskov2022}.

However, a dressed anomaly leaves the structure intact but not the derivation:
the cancellation of Eq.~(\ref{eq:firstorder}) takes $\ano$ outside the
integral, so a phase-dependent $\ano(\chi(\eta))$ would leave
$\int\ano(\eta)\,\nhat\wedge d\aperp\neq0$ and restore a first-order
term~\cite{Ilderton2020,Torgrimsson2021}. At $\chi\simeq2\times10^{-4}$ the
dressing is far too weak to matter, but the statement is exact only for
constant $\ano$, and outside the locally constant field approximation a local
anomaly is not defined at all~\cite{DiPiazzaPatuleanu2021}.

Two uses survive the negative budget of Sec.~\ref{sec:budget}. The identical
zero at $g=2$, and the identical zero at any $g$ for linear polarization, give
structure-preserving spin
integrators~\cite{Hairer2006,SanzSerna1992} a reference that needs no converged
solution to compare against, and the tolerance scan of Sec.~S4.3 shows what
such a test looks like in practice. The classical result is also a well-posed
question at the quantum level, where the exact Dirac--Pauli solutions in a
plane wave~\cite{Chakrabarti1968,Barducci2008,Bagrov2014} make it tractable. We
expect an $O(\ano^{2})$ geometric term in the spin-flip amplitude of
Ref.~\cite{Ilderton2020} at the order beyond the one written there, with a
coefficient predicted by Eq.~(\ref{eq:holonomy}) and no free parameter.

\begin{acknowledgments}
Earlier stages of this work were presented at the 26th International Symposium
on Spin Physics (SPIN 2025), Qingdao, China, 21--26 September 2025, and at the
1st Conference on Strong-Interaction Spin Physics, Qingdao, China, 26--31 July
2026.

This work was partially supported by the State Assignment of the Ministry of
Education and Science of the Russian Federation (Project No.\ FZEN 2023-0006),
the Nantong Science and Technology Plan Project (Grant Nos.\ JC2020137 and
JC2020138), the Key Research and Development Program of Jiangsu Province of
China (Grant No.\ BE2021013-1), the National Natural Science Foundation of
Jiangsu Province of China (Grant No.\ BK20201438), and in part by the Natural
Science Research Project of Jiangsu Provincial Institutions of Higher
Education (Grant Nos.\ 1120KJA510002 and 20KJB510010).
\end{acknowledgments}

N.~S.~A. and A.~P.~N. developed the analytic reduction and wrote the
manuscript. A.~P.~N. performed the plane-wave computations. S.~N.~A.
performed the focused-beam computations and the Maxwell residual tests.
Q.-H.~Q. supervised the project. All authors discussed the results and
approved the final manuscript.

The authors declare no competing financial interest.

\section*{Data availability}
The code that generates the data and the figures of this paper, together with
the computed results themselves, is openly available in the Zenodo repository
at Ref.~\cite{Akintsov2026data}.

%% ===========================================================================
%%  APPENDIX -- the Supplemental Material of the journal version, in full.
%%
%%  The counters are reset and given the "S" prefix that supplement.tex gives
%%  them when it is compiled on its own, so that every internal reference in
%%  the material below, and every hard-coded "Sec. S4.3" in the article above,
%%  prints exactly what it printed in the two separate documents.  REVTeX
%%  prefixes a \ref to a subsection with the section number, which is right
%%  above (where \thesubsection is a bare letter) and wrong below (where
%%  \thesubsection already carries it); \p@subsection drops the prefix.
%% ===========================================================================
\clearpage

\setcounter{section}{0}
\setcounter{equation}{0}
\setcounter{figure}{0}
\setcounter{table}{0}
\renewcommand{\theequation}{S\arabic{equation}}
\renewcommand{\thefigure}{S\arabic{figure}}
\renewcommand{\thetable}{S\arabic{table}}
\renewcommand{\thesection}{S\arabic{section}}
\renewcommand{\thesubsection}{S\arabic{section}.\arabic{subsection}}
\makeatletter
\renewcommand{\p@subsection}{}
\makeatother

%% Resetting the counters makes the appendix start again at 1, so the PDF
%% anchors hyperref builds from the raw counter values would collide with
%% those of the article.  These give the appendix anchors of its own.
\renewcommand{\theHequation}{S\arabic{equation}}
\renewcommand{\theHfigure}{S\arabic{figure}}
\renewcommand{\theHtable}{S\arabic{table}}
\renewcommand{\theHsection}{S\arabic{section}}
\renewcommand{\theHsubsection}{S\arabic{section}.\arabic{subsection}}

\onecolumngrid
\begin{center}
  \rule{0.85\textwidth}{0.4pt}\\[1.6ex]
  {\large\bfseries Appendix: Supplemental Material}\\[1.2ex]
  \begin{minipage}{0.85\textwidth}
    \small
    In the journal version this material is a separate file. It is reproduced
    here unchanged so that the preprint stands on its own. Sections,
    equations, figures and tables below carry the prefix ``S''; ``the main
    text'' means Secs.~\ref{sec:intro}--\ref{sec:outlook} above, whose
    equation numbers carry no prefix.
  \end{minipage}\\[1.2ex]
  \rule{0.85\textwidth}{0.4pt}
\end{center}
\twocolumngrid

\noindent
Sections~\ref{sec:S1} and~\ref{sec:S2} carry the derivation of the holonomy
law in the detail the main text leaves out, from the assumptions behind the
Thomas--Bargmann--Michel--Telegdi (BMT) equation to the non-Abelian Stokes
evaluation of the ordered exponential. Section~\ref{sec:S3} collects the
invariance properties that make the result geometric, including the reason the
storage-ring factor $\ano\gamma$ does not apply. Section~\ref{sec:S4} documents
the numerical protocol, the invariant control, the convergence of the null
result in the integrator tolerance and the validation runs, Sec.~\ref{sec:S5}
the focused-beam model and its limitations, Sec.~\ref{sec:S6} the window in
which the classical treatment is meaningful, and Sec.~\ref{sec:S7} the
mechanisms that break the theorem. Equations, figures, and tables carry the
prefix ``S''. Equation numbers without it refer to the main text. The anomaly
is written $\ano=(g-2)/2$ here and in the main text alike.

Every number quoted below is labeled either \emph{measured}, meaning it was
produced by one of the runs described in Secs.~\ref{sec:S4} and~\ref{sec:S5},
or \emph{estimated}, meaning it follows from a published scaling with no
integration of our own. Nothing in between is reported, and where a quantity
was neither measured nor estimated we say so: Sec.~\ref{sec:S3.4} is the one
place where that happens.

\section{The Thomas--BMT equation in light-front variables}
\label{sec:S1}

\subsection{Field and orbit}

A null wave vector $k^{\mu}$ with $k^{2}=0$ fixes the light-front phase
$\eta=k\cdot x$ in the metric $(+,-,-,-)$, and the dimensionless potential
\begin{equation}
  a^{\mu}(\eta)=\frac{eA^{\mu}(\eta)}{m},
  \qquad k\cdot a=0 ,
  \label{eq:Spot}
\end{equation}
depends on $\eta$ alone. The anomaly is written $\ano=(g-2)/2$, which leaves the
symbol $a$ for the potential and its components alone. The amplitude parameter
is $a_{0}$: in the convention fixed below the peak of $|\bm{a}_{\perp}|$ is
$a_{0}/\sqrt{1+\delta^{2}}$ at ellipticity $\delta$, which is $a_{0}$ for
linear polarization and $a_{0}/\sqrt{2}$ for circular polarization, the
normalization holding the mean square rather than the peak fixed. The vector
$\bm{a}_{\perp}$ carries components $(a_{1},a_{2})=(a_{x},a_{y})$ along a
right-handed polarization basis $\hat{e}_{1}\times\hat{e}_{2}=\nhat$.

Every sign below is referred to a single fixed convention, and all of them have
been rechecked against it. The wedge product of two four-vectors is the mixed
tensor $(p\wedge q)^{\mu}{}_{\nu}=p^{\mu}q_{\nu}-q^{\mu}p_{\nu}$, which acts on
a four-vector $S$ as $p\,(q\cdot S)-q\,(p\cdot S)$. The symbol $\wedge$ never
denotes a three-dimensional wedge below. The propagation direction is
written $\nhat$ both as a unit spatial vector and as the four-vector
$\nhat^{\mu}=\tfrac12(k-\bar{k})^{\mu}$ that represents it in the rest frame of
$u_{0}$, and the same holds for $\bm{a}_{\perp}$, whose four-vector form is
$a^{\mu}$. A bold two-vector and its four-vector are interchangeable inside a
wedge, where only the transverse components contribute. One consequence is
used repeatedly: for spatial $p$ and $q$ and a rest-frame spin $\bm{\zeta}$ the
lowered index flips a sign, $(p\wedge q)\bm{\zeta}=(\bm{p}\times\bm{q})\times
\bm{\zeta}$, so that $e_{1}\wedge e_{2}$ generates a right-handed rotation
about $\nhat$ and $\nhat\wedge\bm{a}_{\perp}'$ a right-handed rotation about
$\nhat\times\bm{a}_{\perp}'$.

Finiteness of the pulse enters once, through
$\bm{a}_{\perp}(+\infty)=\bm{a}_{\perp}(-\infty)=0$, which is Eq.~(1) of the
main text.
The phase average of $\bm{a}_{\perp}$ is not constrained, and
Sec.~\ref{sec:S3.5} returns to that point.

The scans of Secs.~\ref{sec:S4} and~\ref{sec:S5} use
\begin{equation}
  a_{x}=\frac{a_{0}f(\eta)}{\sqrt{1+\delta^{2}}}\cos(\eta+\varphi_{0}),
  \quad
  a_{y}=\frac{a_{0}\delta f(\eta)}{\sqrt{1+\delta^{2}}}\sin(\eta+\varphi_{0}),
  \label{eq:Senvelope}
\end{equation}
with ellipticity $\delta$ ($\delta=0$ linear, $\delta=1$ circular) and
carrier-envelope phase (CEP) $\varphi_{0}$. Two envelopes are used: a Gaussian
$f=\exp(-\eta^{2}/2\sigma^{2})$ with $\sigma=\pi N/\sqrt{\ln 2}=3.7734\,N$,
truncated at $|\eta|\le 8\sigma$, and $f=\cos^{2}(\pi\eta/2L)$ with
$L=8.6294\,N$ on compact support. In both, $N$ is the number of optical cycles
in the full width at half maximum of the intensity envelope $f^{2}$. The two
are compared at equal $N$. The signed area then evaluates to
\begin{equation}
  \Anet=\int\bigl(a_{x}a_{y}'-a_{y}a_{x}'\bigr)\,d\eta
       =\frac{\delta}{1+\delta^{2}}\,a_{0}^{2}\!\int\! f^{2}\,d\eta ,
  \label{eq:Sarea}
\end{equation}
with $\int f^{2}d\eta=\sigma\sqrt{\pi}$ for the Gaussian and $0.75L$ for
$\cos^{2}$.

Because $k_{\mu}F^{\mu\nu}=0$ for a null plane wave, the light-front
projection $\kappa=k\cdot u$ is conserved, $d/d\tau=\kappa\,d/d\eta$, and the
orbit is the Volkov solution~\cite{Volkov1935}
\begin{equation}
  u^{\mu}(\eta)=u_{0}^{\mu}-a^{\mu}
    +\frac{k^{\mu}}{\kappa}\Bigl(a\cdot u_{0}-\tfrac12\,a\cdot a\Bigr),
  \qquad u(\pm\infty)=u_{0}.
  \label{eq:Svolkov}
\end{equation}
In components, $u_{\perp}(\eta)=u_{0\perp}-\bm{a}_{\perp}(\eta)$ and
$u^{0,3}=\tfrac12[\kappa\pm(1+u_{\perp}^{2})/\kappa]$, where
$\kappa=u^{0}-u^{3}$ is the same conserved quantity, written for $k$
normalized to unit frequency. For a head-on collision
$\kappa=\gamma(1+\beta)\simeq2\gamma$. No integration is needed for the
orbit, which removes one source of numerical error from everything that
follows.

\subsection{Assumptions}
\label{sec:S1.2}

The spin is a classical four-vector $S^{\mu}$ with $S\cdot u=0$ and
$S^{2}=-1$, transported by the covariant BMT
equation~\cite{Thomas1926,Bargmann1959,Jackson1999}, Eq.~(2) of the
main text,
\begin{equation}
  \frac{dS^{\mu}}{d\tau}
  = \frac{e}{m}\left[\frac{g}{2}F^{\mu\nu}S_{\nu}
    + \left(\frac{g}{2}-1\right)u^{\mu}
      \bigl(S_{\lambda}F^{\lambda\nu}u_{\nu}\bigr)\right].
  \label{eq:Sbmt}
\end{equation}
Its laboratory form is $d\bm{S}/dt=\bm{\Omega}\times\bm{S}$ with
\begin{equation}
  \bm{\Omega}=-\frac{e}{m}\Bigl[\Bigl(\ano+\tfrac1\gamma\Bigr)\bm{B}
    -\frac{\ano\gamma}{\gamma+1}(\bm{\beta}\!\cdot\!\bm{B})\bm{\beta}
    -\Bigl(\ano+\tfrac1{\gamma+1}\Bigr)\bm{\beta}\times\bm{E}\Bigr],
  \label{eq:Slab}
\end{equation}
and the coefficients of Eq.~(\ref{eq:Slab}) were used to check those of
Eq.~(\ref{eq:Sbmt}) term by term. At $\gamma=1$ the first coefficient is
$\ano+1=g/2$, as it must be for an electron at rest.

The Stern--Gerlach force is dropped: relative to the Lorentz force it carries a
factor $\hbar k/mc$, which is $\hbar\omega/mc^{2}\simeq3\times10^{-6}$ for
Ti:sapphire light in the rest frame of a slow electron, so that at the accuracy
of interest here the orbit is insensitive to the spin. Radiation reaction goes
with it, bounded from published scalings in Sec.~\ref{sec:S6} and never
integrated, and so does photon emission, which is the same statement at the
quantum level. An electric dipole moment is set to zero. The Stern--Gerlach
term is the mildest of the four, by the factor just quoted. The other three
are genuine restrictions.

Within the plane-wave sector the classical transport law is not itself an
approximation, which limits the damage.
Chakrabarti solved the Dirac--Pauli equation exactly for fields of the form
$F=F(k\cdot x)$ with $k^{2}=0$ and showed that the polarization four-vector
obeys the BMT equation exactly, with an arbitrary anomalous magnetic moment
and an arbitrary electric dipole moment~\cite{Chakrabarti1968}. Barducci and
Giachetti later constructed the eigenfunctions and the Green function of the
same problem~\cite{Barducci2008}. The classical spin transport used below is
therefore the exact one-particle content of the Dirac--Pauli theory in this
background, and our treatment omits radiation while keeping the spin
kinematics exact.

\subsection{A single antisymmetric generator}

Since $u\cdot S=0$, the right-hand side of Eq.~(\ref{eq:Sbmt}) can be brought
to antisymmetric form. The completion is unique: two antisymmetric matrices
that act identically on the three-dimensional space orthogonal to $u$ differ
by an antisymmetric matrix with a three-dimensional kernel. The rank of an
antisymmetric matrix is even. That difference therefore vanishes. Writing
$f=(e/m)F=k\wedge a'$ and $w^{\mu}=f^{\mu\nu}u_{\nu}$,
\begin{equation}
  \frac{dS^{\mu}}{d\eta}=\frac{1}{\kappa}\,\Omega^{\mu}{}_{\nu}S^{\nu},
  \qquad
  \Omega=(1+\ano)\,k\wedge a' + \ano\,(u\wedge w) ,
  \label{eq:Sgenerator}
\end{equation}
which is Eq.~(3) of the main text.

The rest of Secs.~\ref{sec:S1} and~\ref{sec:S2} is written for a head-on
geometry normalized to $\kappa=1$, and the restriction is lifted here, one
paragraph before it is imposed. A null rotation $N$ in the little group
$ISO(2)_{k}$ of $k$ leaves $k$ fixed and sends $e_{i}\mapsto e_{i}+c_{i}k$, so
that $NfN^{-1}=k\wedge(a'+ck)=f$: the background is invariant under
$ISO(2)_{k}$, which is stronger than gauge equivalence, and since
$ISO(2)_{k}$ acts transitively on the mass shell at fixed $\kappa$, any
$u_{0}$ can be carried to $(k+\bar{k})/2$. The net map of the original geometry
is then the conjugate $N^{-1}\Lambda_{\rm net}N$, and a rotation angle is a
conjugation invariant.
This matters because Eq.~(\ref{eq:Suwedgew}) below, and with it the reduction
of Sec.~\ref{sec:S2.2} and Eq.~(7) of the main text, is false as written when
$u_{0\perp}\neq0$: the general form of the reduced generator carries $\kappa$,
$M=\tfrac12(k/\kappa-\kappa\bar{k})\wedge a'$, in which $\kappa$ acts as a
rescaling of $\eta$ and nothing else. Section~\ref{sec:S3.2} gives the measured
values.

Expanding $u\wedge w$ in the basis $\{k,\bar{k},e_{1},e_{2}\}$ with
$k\cdot\bar{k}=2$ and $e_{i}\cdot e_{j}=-\delta_{ij}$, and with
$u_{0}=(k+\bar{k})/2$,
\begin{multline}
  u\wedge w=\tfrac{\mathcal{U}}{2}\,(\bar{k}\wedge k)
    -\tfrac12\,\bar{k}\wedge a'\\
    -\Bigl(\tfrac12+\tfrac{|\bm{a}_{\perp}|^{2}}{2}\Bigr)k\wedge a'
    +\mathcal{U}\,k\wedge a
    +\mathcal{W}\,(e_{1}\wedge e_{2}) ,
  \label{eq:Suwedgew}
\end{multline}
with
\begin{equation}
  \mathcal{U}=\bm{a}_{\perp}\!\cdot\!\bm{a}_{\perp}' ,
  \qquad
  \mathcal{W}=a_{1}a_{2}'-a_{2}a_{1}' .
  \label{eq:SUW}
\end{equation}
The symbol $\chi$ is reserved throughout for the quantum nonlinearity parameter
of Sec.~\ref{sec:S6}.

However, two of the five terms in Eq.~(\ref{eq:Suwedgew}) lie outside the
little algebra $\mathfrak{iso}(2)$ of $k$: the boost $\bar{k}\wedge k$ and the
conjugate null rotation $\bar{k}\wedge a'$. Together with $k\wedge e_{i}$ they
generate a nonabelian $\mathfrak{sl}(2,\mathbb{R})$ already for linear
polarization. Hence the $\eta$ ordering in Eq.~(\ref{eq:Sgenerator}) is not
degenerate at $g\neq2$, and the one-line argument of Sec.~\ref{sec:S2.1} does
not extend to it.

\section{Derivation of the holonomy law}
\label{sec:S2}

\subsection{The $g=2$ propagator}
\label{sec:S2.1}

At $\ano=0$ the generator is $\Omega=k\wedge a'=a_{1}'N_{1}+a_{2}'N_{2}$ with
$N_{i}=k\wedge e_{i}$. Direct multiplication gives
\begin{equation}
  (N_{i}N_{j})^{\mu}{}_{\nu}=\delta_{ij}\,k^{\mu}k_{\nu} ,
  \label{eq:SNiNj}
\end{equation}
which is symmetric in $(ij)$, so $[N_{1},N_{2}]=0$ and $N_{i}^{3}=0$. The
generator commutes with itself at different phases, the ordering collapses,
and
\begin{equation}
  \Lambda_{0}(\eta)=\exp\bigl[a_{1}(\eta)N_{1}+a_{2}(\eta)N_{2}\bigr],
  \qquad \Lambda_{0}(+\infty)=\mathbb{1},
  \label{eq:Slambda0}
\end{equation}
by Eq.~(1) of the main text.
No property of $u_{0}$ enters, since $u$ does not appear in $\Omega$ at $\ano=0$
at all. This is the statement that the asymptotic spin index of a Volkov state
is conserved~\cite{Volkov1935}, and for linear polarization it is the
vanishing of the electric pulse area found by Aleksandrov \textit{et
al.}~\cite{Aleksandrov2020}.

\subsection{Interaction picture and the exact reduction}
\label{sec:S2.2}

Since $(1+\ano)f+\ano\,u\wedge w=f+\ano\,(f+u\wedge w)$, factoring the
evolution as
$U=\Lambda_{0}V$ gives
\begin{equation}
  \frac{dV}{d\eta}=\ano\,M(\eta)\,V ,
  \qquad
  M={\rm Ad}_{\Lambda_{0}^{-1}}\bigl(f+u\wedge w\bigr).
  \label{eq:SVeq}
\end{equation}
Write $A=k\wedge\bm{a}_{\perp}=a_{1}N_{1}+a_{2}N_{2}$, so that
$\Lambda_{0}=e^{A}$ and $M=e^{-{\rm ad}_{A}}X$ with $X=f+u\wedge w$. The series
terminates after three terms. Nilpotency of ${\rm ad}_{A}$ is what stops it,
and on the whole of $\mathfrak{so}(1,3)$ the operator $A$ raises the grading
$\bar{k}\wedge e_{i}\to\{\bar{k}\wedge k,\,e_{1}\wedge e_{2}\}\to
k\wedge e_{i}\to0$, so ${\rm ad}_{A}^{3}=0$ and
$M=X-[A,X]+\tfrac12[A,[A,X]]$. Nilpotency of the $N_{i}$ alone would give
only ${\rm ad}_{A}^{5}=0$.

\begin{table*}[!tbp]
  \caption{The three surviving terms of $M=e^{-{\rm ad}_{A}}X$ with
    $X=f+u\wedge w$, decomposed in the bivector basis of $\mathfrak{so}(1,3)$
    and evaluated at $\bm{a}_{\perp}=(0.7,-1.3)$,
    $\bm{a}_{\perp}'=(0.4,0.9)$ for the head-on geometry at $\kappa=1$. The
    entries are exact algebra evaluated at one point. The two columns
    outside the polarization plane, $\bar{k}\wedge k$ and $e_{1}\wedge e_{2}$,
    are emptied by the first commutator alone; the identities of
    Eq.~(\ref{eq:Sidentities}) act only on the two $k\wedge e_{i}$ columns. The
    sum reproduces $\tfrac12(k-\bar{k})\wedge a'=\nhat\wedge a'$ with
    $a'=(0.4,0.9)$.}
  \label{tab:Sadjoint}
  \squeezetable
  \footnotesize
  \begin{ruledtabular}
  \begin{tabular}{lrrrrrr}
    & $k\wedge e_{1}$ & $k\wedge e_{2}$ & $\bar{k}\wedge e_{1}$
      & $\bar{k}\wedge e_{2}$ & $\bar{k}\wedge k$ & $e_{1}\wedge e_{2}$ \\
    \hline
    $X$ & $-0.859$ & $0.626$ & $-0.2$ & $-0.45$ & $-0.445$ & $1.15$ \\
    $-[A,X]$ & $2.118$ & $-0.352$ & $0$ & $0$ & $0.445$ & $-1.15$ \\
    $\tfrac12[A,[A,X]]$ & $-1.059$ & $0.176$ & $0$ & $0$ & $0$ & $0$ \\
    $M$ & $0.2$ & $0.45$ & $-0.2$ & $-0.45$ & $0$ & $0$ \\
  \end{tabular}
  \end{ruledtabular}
\end{table*}

Table~\ref{tab:Sadjoint} carries the three terms out at one numerical point and
shows where each cancellation happens. Everything along the boost
$\bar{k}\wedge k$ and along $e_{1}\wedge e_{2}$ is removed by the first
commutator, with no input beyond the algebra. The two algebraic identities
\begin{equation}
  \mathcal{U} a_{1}-\mathcal{W}a_{2}=|\bm{a}_{\perp}|^{2}a_{1}' ,
  \qquad
  \mathcal{U} a_{2}+\mathcal{W}a_{1}=|\bm{a}_{\perp}|^{2}a_{2}' ,
  \label{eq:Sidentities}
\end{equation}
which follow from Eq.~(\ref{eq:SUW}) by inspection, do something else: they
collapse the residual null rotations, a combination of $k\wedge a'$,
$k\wedge\bm{a}_{\perp}$ and $k\wedge\tilde{\bm{a}}$ with
$\tilde{\bm{a}}=a_{2}e_{1}-a_{1}e_{2}$, into the single term
$\tfrac12\,k\wedge a'$. What survives is
\begin{equation}
  M(\eta)=\tfrac12(k-\bar{k})\wedge a'(\eta)=\nhat\wedge a'(\eta) ,
  \label{eq:SM}
\end{equation}
equivalently, for the rest-frame spin $\bm{\zeta}$ and in the convention of
Sec.~\ref{sec:S1},
\begin{equation}
  \frac{d\bm{\zeta}}{d\eta}
  = \ano\,\bigl(\nhat\times\bm{a}_{\perp}'\bigr)\times\bm{\zeta} ,
  \label{eq:Szeta}
\end{equation}
which is Eq.~(7) of the main text.
We evaluated the components of Eq.~(\ref{eq:Sgenerator}) on a grid in $\eta$
and compared them with Eq.~(\ref{eq:SM}). The largest discrepancy is
$2.7\times10^{-15}$ (measured).

$M$ is a pure rotation of the rest frame about an axis
$\nhat\times\bm{a}_{\perp}'$ lying in the polarization plane. For linear
polarization that axis is fixed, the rotations at different phases commute, and
$V(+\infty)=\exp[\ano\,\Delta a_{1}\,\nhat\wedge e_{1}]=\mathbb{1}$ exactly at
any
$g$ and any $a_{0}$. For $u_{0\perp}\neq0$ the conjugation of
Sec.~\ref{sec:S1} carries the conclusion over unchanged, since
$N^{-1}\mathbb{1}N=\mathbb{1}$. This is the mechanism behind the vanishing
quoted in the main text. It is perturbative in nothing.

Cancellation of the first order in the anomaly holds for every polarization,
and its cause is the one that also makes the whole construction geometric.
$M\,d\eta=\nhat\wedge d\bm{a}_{\perp}$ is a one-form on the polarization plane
whose coefficients are constants, so its integral along a path depends on the
endpoints alone, and Eq.~(1) of the main text says that the two endpoints
coincide. Hence $\ano\int M\,d\eta=\ano\,\nhat\wedge\Delta\bm{a}_{\perp}=0$,
exactly,
at any $a_{0}$ and for any curve. A weaker fact is sometimes offered in place
of this one and does not do the work. $\Lambda_{\rm net}$ lies in the
stabilizer $SO(3)_{u_{0}}$ of $u_{0}$, which follows from $u(\pm\infty)=u_{0}$
and the conservation of $S\cdot u$, but membership in $SO(3)_{u_{0}}$ confines
the first-order coefficient to a three-dimensional space and constrains it no
further. Section~\ref{sec:S5.4} makes that concrete, since in a focused beam
the net map still lies in $SO(3)$ once the boost is subtracted, while its
first-order coefficient $r_{1}$ is nonzero and grows as $\varepsilon^{2}$.

The geometric statement can now be given in full. The base is the polarization
plane $\mathbb{R}^{2}$ with coordinates $\bm{a}_{\perp}=(a_{1},a_{2})$, the
bundle is the trivial principal bundle $\mathbb{R}^{2}\times SO(3)_{u_{0}}$
whose structure group is the stabilizer of $u_{0}$ in $SO(1,3)$, and $M$ lies
in the corresponding Lie algebra because $Mu_{0}=0$. The connection one-form is
\begin{equation}
  \omega=\ano\,\bigl(G_{1}\,da_{1}+G_{2}\,da_{2}\bigr) ,
  \qquad
  G_{i}=\nhat\wedge e_{i} ,
  \label{eq:Sconnection}
\end{equation}
with coefficients independent of $\eta$ and of $\bm{a}_{\perp}$ alike. The path
is $\mathcal{C}:\eta\mapsto\bm{a}_{\perp}(\eta)$, and it closes for exactly one
reason, Eq.~(1) of the main text. That is the only place where finiteness of
the
pulse is used. The net map is the holonomy of $\omega$ around that loop,
\begin{equation}
  \Lambda_{\rm net}
  =\mathrm{Hol}_{\omega}(\mathcal{C})
  =\mathcal{P}\exp\oint_{\mathcal{C}}\omega ,
  \label{eq:Sholonomy}
\end{equation}
and it depends on $\mathcal{C}$ alone, with no reference to the
parameterization. No adiabatic parameter appears here and no eigenspace is
being followed, so this is ordinary
non-Abelian parallel transport, and the resemblance to the geometric phases of
Refs.~\cite{Berry1984,Wilczek1984} is an analogy. The one-form is also exact,
but exactness is not what does the work, since in the non-Abelian case an exact
form can still have nontrivial holonomy and reparameterization invariance holds
for any one-form whatsoever. Constant coefficients are what matter.

\subsection{Non-Abelian Stokes evaluation and the order in the anomaly}
\label{sec:S2.3}

The curvature of $\omega$ is constant over the plane,
\begin{equation}
  [G_{1},G_{2}]=e_{1}\wedge e_{2},
  \quad
  F=d\omega+\omega\wedge\omega=\ano^{2}(e_{1}\wedge e_{2})\,da_{1}\wedge da_{2},
  \label{eq:Scurvature}
\end{equation}
and $e_{1}\wedge e_{2}$ generates a right-handed rotation about $\nhat$. The
version of the non-Abelian Stokes theorem that applies is the surface-ordered
one~\cite{Halpern1979,Broda2001}, in which the path-ordered exponential around
$\mathcal{C}$ becomes a surface-ordered exponential of the curvature carried
back to a base point along a mesh of paths. Constancy of $F$ in the coordinates
of the base does not make that ordering trivial, because what is ordered is not
$F$ but its adjoint transport along the mesh, and that is not constant. The
theorem therefore delivers the leading term and nothing more: a rotation about
$\nhat$ through $\ano^{2}$ times the oriented area enclosed by $\mathcal{C}$,
with
the overall sign, which turns on the orientation convention relating the
boundary to the surface ordering, taken below from the Magnus expansion. One
further caution belongs here. The area is algebraic and counted with
multiplicity, so that for an $N$-cycle envelope, where $\mathcal{C}$ winds
around the origin roughly $N$ times, $\Anet$ counts $2N$ times the area of one
turn.

The second Magnus term~\cite{Magnus1954,Blanes2009} gives the same leading
result in three lines, and it fixes the sign. The rotation vector of
Eq.~(\ref{eq:Szeta}) is $\bm{m}(\eta)=\ano\,\nhat\times\bm{a}_{\perp}'(\eta)$,
which is $\ano\bm{a}_{\perp}'$ turned through a right angle about $\nhat$.
Turning
both arguments of a cross product of planar vectors through the same right
angle leaves its $\nhat$ component alone, so the double integral may be written
with $\bm{a}_{\perp}'$ directly,
\begin{equation}
  \Theta_{\rm net}
  =\frac{\ano^{2}}{2}\iint_{\eta_{1}<\eta_{2}}
    \bigl[\bm{a}_{\perp}'(\eta_{2})\times\bm{a}_{\perp}'(\eta_{1})\bigr]
    \cdot\nhat\;d\eta_{1}d\eta_{2} .
  \label{eq:Smagnus}
\end{equation}
Integrating the inner variable and using $\bm{a}_{\perp}(-\infty)=0$ converts
the double integral into $-\int(a_{x}a_{y}'-a_{y}a_{x}')\,d\eta$, so
\begin{equation}
  \Theta_{\rm net}=-\tfrac{1}{2}\ano^{2}\Anet+O(\ano^{4}) ,
  \label{eq:Sresult}
\end{equation}
in the triad $\hat{e}_{1}\times\hat{e}_{2}=\nhat$. This is Eq.~(10) of the
main text,
and $\Anet$ is twice the algebraic area enclosed by $\mathcal{C}$.

The order of the remainder is a theorem, and the proof takes two lines.
Grade $\mathfrak{so}(3)_{u_{0}}$ by calling $G_{1}$ and
$G_{2}$ odd and $e_{1}\wedge e_{2}$ even. The bracket respects that grading,
since $[{\rm odd},{\rm odd}]={\rm even}$ and
$[{\rm even},{\rm odd}]={\rm odd}$. Every term of the Magnus series is an
integral of $n$-fold nested brackets of $M$, and $M$ is odd, so the $n$th term
has parity $n$: terms with $n$ even lie along $e_{1}\wedge e_{2}$, and terms
with $n$ odd lie in the plane spanned by $G_{1}$ and $G_{2}$. The consequences
follow immediately. The component of the exponent along $\nhat$ carries
only even powers of $\ano$, which establishes the $O(\ano^{4})$ in
Eq.~(\ref{eq:Sresult}); the leading correction to the axis is the
third-order in-plane term, so at fixed curve the axis is tilted out of $\nhat$
by $O(\ano)$; and the change in the projection of the spin on $\nhat$ is
quadratic
in that tilt and therefore $O(\ano^{6})$. Section~\ref{sec:S3.4} uses the last
of
the three. Without it that statement would rest on nothing.

Equation~(\ref{eq:Sresum}) below sharpens the remainder further, since the
relative correction there is governed by $\ano a_{0}$, so the informative form
is
$\Theta_{\rm net}=-\tfrac12\ano^{2}\Anet[1+O((\ano a_{0})^{2})]$. The Magnus
series
itself, on the other hand, has a guaranteed radius of convergence only while
$\ano\int\|M\|\,d\eta=\ano\,{\rm length}(\mathcal{C})<\pi$, and the working
point of
Sec.~\ref{sec:S6} violates that bound outright: at $a_{0}=75$ and $N=32$ the
curve has length $\simeq2\pi Na_{0}/\sqrt{2}\simeq1.1\times10^{4}$, so
$\ano\,{\rm length}(\mathcal{C})\simeq12\gg\pi$. Hence nothing quoted at that
point rests on the series. What supports Eq.~(\ref{eq:Sresult}) there is the
resummation of Sec.~\ref{sec:S2.4}, which sums all orders in $\ano a_{0}$ for a
circular plateau, together with the measured ratios of Sec.~\ref{sec:S4.4}.

One definition has to be pinned down before Eq.~(\ref{eq:Sresult}) can be read
literally. Its right-hand side is an unbounded real number growing as
$\ano^{2}a_{0}^{2}N$, whereas the angle extracted from an $SO(3)$ matrix lies in
$[0,\pi]$, is defined modulo $2\pi$, and carries a sign ambiguity that goes
with a reversal of the axis. We define $\Theta_{\rm net}$ as the continuous
lift in $\ano$ from $\ano=0$, equivalently as the angle accumulated along the
path,
and the measured quantity is its reduction modulo $2\pi$. The wrapping is not
hypothetical. For a four-cycle circular plateau with one-cycle fronts at
$a_{0}=3$ and $\ano=0.2$, where $\Anet=268.61$ and $\ano a_{0}=0.6$ is not
small, Eq.~(\ref{eq:Sresum}) predicts $4.96$~rad and the leading order
Eq.~(\ref{eq:Sresult}) gives $5.37$~rad; the angle read from the matrix is
$1.307$~rad, which is $2\pi$ minus $4.976$~rad, and the extracted axis changes
sign (measured; \texttt{code/holonomy\_scan.json}, stage \texttt{wrap}).
Section~\ref{sec:S4.4} drops the affected points from the scaling
table for the same reason.

For linear polarization $\mathcal{C}$ is a segment and $\Anet=0$, and the
vanishing there is exact at every order in $\ano$ and $a_{0}$ by the fixed-axis
argument of Sec.~\ref{sec:S2.2}, well beyond the order of
Eq.~(\ref{eq:Sresult}). For a curve traced out and back, such as
$a_{y}\propto a_{x}^{2}$, the enclosed area is zero and the ordered exponential
is the identity for the same reason. We measured $\le9.95\times10^{-12}$~rad
over the 18 cases of the FIG.~1 grid, with anomalies up to $\ano=0.6$. The
covariant cross-check of Sec.~\ref{sec:S4} returns $\le7.7\times10^{-11}$~rad
on four cases of its own, at its own coarser floor.

\subsection{Resummation of the axis tilt for a circular plateau}
\label{sec:S2.4}

Let $\bm{a}_{\perp}$ trace a circle of radius $a_{0}$ at unit rate in $\eta$
over a phase interval $\Delta\eta$, so that $\Anet=a_{0}^{2}\Delta\eta$. Then
$\nhat\times\bm{a}_{\perp}'$ has fixed magnitude $a_{0}$ and rotates about
$\nhat$ at unit rate, and Eq.~(\ref{eq:Szeta}) becomes the magnetic-resonance
problem for a rotating field of strength $\ano a_{0}$ on exact resonance. In the
frame corotating about $\nhat$ the rotation vector is constant with magnitude
$\sqrt{1+\ano^{2}a_{0}^{2}}$. Transforming back subtracts one turn per unit
phase, and the net rotation about $\nhat$ has magnitude
\begin{equation}
  \Theta_{\rm circ}
  = \Bigl(\sqrt{1+\ano^{2}a_{0}^{2}}-1\Bigr)\,\Delta\eta ,
  \label{eq:Sresum}
\end{equation}
Eq.~(12) of the main text,
with the sense of rotation of Eq.~(\ref{eq:Sresult}). Expanding for
$\ano a_{0}\ll1$ returns
$\tfrac12\ano^{2}a_{0}^{2}\Delta\eta=\tfrac12\ano^{2}\Anet$.
At $a_{0}=1$ and $\ano=0.05$ the integrator gives $2.3193\times10^{-2}$, the
resummation $2.3193\times10^{-2}$, and the area law
$2.3208\times10^{-2}$ (measured).

The word exact does not belong to Eq.~(\ref{eq:Sresum}), and three conditions
say why. An unbroken circle violates
$\bm{a}_{\perp}(\pm\infty)=0$ and is therefore not an admissible pulse, so
Eq.~(\ref{eq:Sresum}) describes a plateau and not a pulse. Composing the
corotating rotation with the compensating turn returns a single angle only for
an integer number of turns; away from that the two axes differ and the result
is a rotation whose angle is not the difference of the two. And for a pulse
with fronts the phase interval has to be read as $\Anet/a_{0}^{2}$, which is
the form printed in the main text; reading it as the plateau length instead
costs $19\%$ at $a_{0}=1$ and $\ano=0.05$. What survives all three is an
asymptotic improvement, not an identity: at $\ano a_{0}=0.3$
Eq.~(\ref{eq:Sresum}) is still off by $0.1\%$. The corresponding statement
about the axis is cleaner, and it is the one the section is named for. In the
corotating frame the rotation vector is tilted from $\nhat$ by
$\arctan(\ano a_{0})$, which is the $O(\ano)$ tilt that the grading argument
of Sec.~\ref{sec:S2.3} predicts, resummed in $\ano a_{0}$.

Note the convention. The radius of the circle in Eq.~(\ref{eq:Sresum}) is
$a_{0}$, whereas in the ellipticity-normalized parameterization of
Eq.~(\ref{eq:Senvelope}) a circularly polarized pulse has radius
$a_{0}/\sqrt{2}$. The two differ by a factor of two in $\Anet$, and the
numbers quoted elsewhere in this document use
Eq.~(\ref{eq:Senvelope}).

The expansion parameter is $\ano a_{0}$, not $\ano$. With the physical anomaly
$\ano a_{0}=1.16\times10^{-3}a_{0}$, so Eq.~(\ref{eq:Sresult}) is accurate to
$10^{-2}$ up to $a_{0}\sim100$. Above $\ano a_{0}\sim1$ the net rotation becomes
linear in $\ano$, which no laboratory field reaches.

\section{Invariance properties}
\label{sec:S3}

\subsection{Reparameterization of the phase}
\label{sec:S3.1}

Equation~(\ref{eq:Sholonomy}) contains $\eta$ only through
$\bm{a}_{\perp}(\eta)$. Under a monotone reparameterization $\eta\to s(\eta)$
that leaves the curve $\mathcal{C}$ unchanged, the one-form
$\nhat\wedge d\bm{a}_{\perp}$ is unchanged and so is $\Lambda_{\rm net}$. The
statement holds by construction, so the numerical test is a control on the
code and not evidence for the theorem: net rotations for one curve traced at
different rates agree to eight or nine digits over eight runs, including at the
inflated value $\ano=0.25$, where the leading order of Eq.~(\ref{eq:Sresult}) is
far from sufficient (measured).

What drops out is the rate, and only the rate, so the carrier frequency, a
chirp and the envelope shape have to be handled with care, because in general
they change $\mathcal{C}$ itself. The correct statement is that everything
about the field enters through the curve $\mathcal{C}$ exactly, and at
leading order through $\Anet$ alone. Lengthening the pulse at
fixed $a_{0}$ enlarges the curve, adds turns and raises $\Anet$, so
$\Theta_{\rm net}\propto N$. Changing the envelope at fixed $N$ and fixed
$a_{0}$ likewise changes $\Anet$, and Sec.~\ref{sec:S4.4} measures by how
much: $\Anet=334.412$ for the Gaussian against $323.603$ for $\cos^{2}$ at
$a_{0}=10$ and $N=1$, a difference of $3.2\%$ that propagates straight into
$\Theta_{\rm net}$. A quadratic chirp on a symmetric envelope leaves $\Anet$
alone by parity, since the chirp-dependent part of $\varphi'$ in
$\Anet\propto\int f^{2}\varphi'\,d\eta$ is odd about the center of an even
$f^{2}$. That is a control on the code and not a law.

\subsection{Independence of the electron energy}
\label{sec:S3.2}

The null rotation generated by $N_{1}=k\wedge e_{1}$ leaves $k$ and $\kappa$
fixed, shifts $u_{0\perp}$ by $-c\kappa\,e_{1}$ for a parameter $c$, and
changes the polarization vectors
only by terms $\propto k^{\mu}$, so that the field tensor itself comes back
unchanged, $NfN^{-1}=f$, which is stronger than gauge equivalence. Choosing
$c=u_{01}/\kappa$ brings any geometry to head-on, the net map is
conjugated, $\Lambda_{\rm net}\to N\Lambda_{\rm net}N^{-1}$, and a rotation
angle is a conjugation invariant. That disposes of the transverse momentum.
The light-front projection $\kappa$ needs a different argument, because
$\kappa=k\cdot u$ is a Lorentz invariant and no boost removes it. It appears
solely through the scale chosen for $k$, hence as a rescaling
$\eta\to\eta/\kappa$ of the phase, to which the holonomy is insensitive by
Sec.~\ref{sec:S3.1}. Thus the inputs that remain are the curve $\mathcal{C}$
and the number $\ano$, and $\gamma$ is not among them.

Measured values agree. For circular polarization at $a_{0}=5$ and $N=1$ the net
rotation is $5.621424\times10^{-5}$~rad at $\gamma=1,10,10^{2},10^{3}$ and
$5.621423\times10^{-5}$ at $\gamma=10^{4}$, six matching digits over four
decades in energy. With the anomaly switched on, an initial transverse momentum
up to $u_{0\perp}=17$ leaves the angle at $1.14759016\times10^{-5}$, and a
copropagating geometry with $\kappa=0.025$ gives the same. The $g=2$ null test
of the main text was run to $u_{0\perp}=15$. Only the rotation axis moves, and
it
tracks the propagation direction of the wave in the electron rest frame to four
digits (measured).

\subsection{Why the storage-ring factor $\ano\gamma$ does not apply}
\label{sec:S3.3}

In a storage ring the field is static in the laboratory, and the ratio of the
anomalous precession frequency to the cyclotron frequency is
$\ano\gamma$~\cite{Baier1972,Mane2005,Jackson1999}. That ratio compares
rates at one instant of laboratory time, and the enhancement comes
from the rest-frame field being $\gamma$ times the laboratory field at fixed
$B$. Neither ingredient survives here. The strength of a plane wave in the
rest frame is measured by the invariant $a_{0}$, which no boost changes, and
the $g=2$ part of the precession, which is what $\ano\gamma$ multiplies, gives a
net rotation of exactly zero. The product is $\ano\gamma\cdot0=0$.

This is a question we have been asked more than once, so we state the
practical form: the enhancement is real, but it applies to a quantity that
vanishes identically in this geometry, and the surviving quantity is
$O(\ano^{2})$ with no compensating factor, so that the one configuration in this
paper where $\ano\gamma$ does return is the static field superposed on the
wave in
Sec.~\ref{sec:S7}, where the added field tensor no longer lies in
$k\wedge(\cdot)$ and the argument of Sec.~\ref{sec:S2.1} has nothing to act on.

\subsection{Helicity}
\label{sec:S3.4}

The grading argument of Sec.~\ref{sec:S2.3} settles the structure here.
At fixed curve the rotation axis of $\Lambda_{\rm net}$
departs from $\nhat$ by $O(\ano)$, the projection of the spin on the propagation
direction changes at $O(\ano^{6})$, and the observable is therefore the rotation
of a transverse polarization. That fixes the experimental configuration, since
a longitudinally polarized beam feels nothing at any order within reach.
Section~\ref{sec:S2.4} adds one qualification: the natural expansion parameter
is $\ano a_{0}$, so at large $a_{0}$ the tilt should be read as $O(\ano a_{0})$.

The exponent is measured, and not left as an inference from the grading.
With the spin started along $\nhat$, at $\gamma=1$, $N=1$ and circular
polarization, and with the anomaly inflated so that $\ano a_{0}$ covers
$0.03$ to $0.5$, the transported triad gives
$\Delta S_{\parallel}=R_{33}-1=-1.80\times10^{-3}(\ano a_{0})^{6}$, with local
log-log exponents of $5.98$ to $6.00$ at the small-anomaly end and a
coefficient that repeats to $0.5\%$ over a tenfold range in $a_{0}$. The
mechanism was checked factor by factor: $\Theta_{\rm net}\propto\ano^{2.000}$,
the axis tilt $\psi\propto\ano^{0.996}$, and
$\Delta S_{\parallel}=-(1-\cos\Theta_{\rm net})\sin^{2}\psi$ to
$5\times10^{-3}$ in relative terms. The orthogonality error of the transported
triad is $1.5\times10^{-14}$, two decades below the smallest measured
$\Delta S_{\parallel}$. Two limits are worth naming. The exponent belongs to
the helicity state, since for a spin that does not start along $\nhat$ the
projection on $\nhat$ changes at $O(\ano^{3})$, measured exponent $3.00$. The
coefficient was measured at one pulse length, one Lorentz factor and one
ellipticity, so it does not transfer to the working point of
Sec.~\ref{sec:S6}, where $\Theta_{\rm net}$ is already $0.4$~rad and the fit
window used here has been left behind (\texttt{code/spin\_magnitude.json},
key \texttt{t7}).

The net operator itself does not depend on the initial spin direction, since we
transport a full orthonormal triad, and only the projection onto a chosen
initial direction does.

\subsection{The pulse condition, stated precisely}
\label{sec:S3.5}

Maxwell's equations and finiteness of the pulse require
$\bm{a}_{\perp}(+\infty)=\bm{a}_{\perp}(-\infty)$, which is the absence of a dc
component in the transverse electric field. They do not require
$\int \bm{a}_{\perp}\,d\eta=0$, which is a condition on a net displacement.
Forcing the second condition by subtracting a pedestal $c_{f}f(\eta)$ with
$c_{f}=e^{-\sigma^{2}/2}\cos\varphi_{0}$ changes the net rotation by
$2.3\times10^{-5}$ in relative terms at $a_{0}=10$, $N=1$, $\delta=1$
($2.248550\times10^{-4}$ against $2.248498\times10^{-4}$, measured). A pulse
deliberately built with $\int \bm{a}_{\perp}d\eta\neq0$ obeys
Eq.~(\ref{eq:Sresult}) to $10^{-5}$ in relative terms at $\Anet=-0.14159$.

A pulse that violates the first condition is a different matter. It breaks the
theorem at $g=2$, and Sec.~\ref{sec:S7} takes that case up.

\subsection{Carrier-envelope phase}
\label{sec:S3.6}

For strictly circular polarization a CEP shift is a rigid rotation of
$\mathcal{C}$ about the origin. The connection of Eq.~(\ref{eq:SM}) is
equivariant under rotations about $\nhat$, and the holonomy angle is a class
function, so the CEP drops out at every order in $\ano$, and the ten matching
digits measured at $\ano=0.25$ are again a control on the code. For elliptical
polarization a CEP-odd part appears only in higher orders of the holonomy:
$8\times10^{-5}$ in relative terms at $\ano=0.25$,
which extrapolates to $\sim2\times10^{-9}$ relative and $\sim10^{-14}$~rad
absolute at the physical anomaly. The largest CEP-dependent part seen anywhere
in the plane-wave scans is $2.4\times10^{-10}$~rad, at $\delta=0$ and
$\gamma=5000$, consistent with the energy-dependent floor of
Sec.~\ref{sec:S4.2} at that $\gamma$.

\section{Numerical verification}
\label{sec:S4}

\subsection{Method}

Two independent implementations were used. The first
(\texttt{code/spin\_magnitude.py}) works in the laboratory frame, writes the
BMT equation componentwise in the form
\begin{multline}
  \frac{dS^{\mu}}{d\eta}
    =(1+\ano)\bigl[n^{\mu}(a'\!\cdot\!S)-a'^{\mu}(n\cdot S)\bigr]\\
    +\,\ano\,u^{\mu}\bigl[(n\cdot S)(a'\!\cdot\!u)-(a'\!\cdot\!S)\bigr],
  \label{eq:Slabform}
\end{multline}
with $n^{\mu}=k^{\mu}/(k\cdot u_{0})$, and evaluates the orbit from
Eq.~(\ref{eq:Svolkov}) analytically at every step. The second
(\texttt{code/indep\_check*.py}) assembles the generator from covariant
bivectors in a different coordinate system, with the wave along $-z$ and the
electron initially at rest. The two were written from the same equations but
not from the same code, and they agree on every conclusion drawn below. They do
not agree on how small a zero is: the second implementation reads the angle
from the trace of the rotation matrix, an operation that is badly conditioned
near the identity, and its floor is four to seven orders of magnitude coarser,
with $\|R-\mathbb{1}\|$ reaching $2.1\times10^{-9}$ and the extracted angle
reaching $6.6\times10^{-6}$~rad on runs whose exact answer is zero. Every
bound quoted below, and in particular the $g=2$ bound of
Table~\ref{tab:Svalid}, comes from the first implementation. The covariant
floor appears twice, in Sec.~\ref{sec:S2.3} and in the last column of
Table~\ref{tab:Svalid}, and is named as such in both places.

Both integrate with the DOP853 explicit Runge--Kutta scheme, the first at
${\rm rtol}=10^{-13}$ and ${\rm atol}=10^{-16}$, the second at
${\rm rtol}=10^{-12}$ and ${\rm atol}=10^{-14}$. Three orthonormal rest-frame
four-vectors are transported together with the orbit. At the end the reference
triad is rebuilt from the initial one, and the net rotation is read off the
resulting $3\times3$ matrix as a rotation vector. Reading the angle from a
transported frame is what makes the result independent of the initial
polarization.

An earlier version of the plane-wave code integrated the four-velocity along
with the spin. That version wasted about half its steps enforcing an identity
that Eq.~(\ref{eq:Svolkov}) already satisfies exactly, and it lost two digits
at $\gamma=10^{3}$. The runs reported here use the closed-form orbit.

\subsection{Invariants and the resolution floor}
\label{sec:S4.2}

The exact invariants are $u\cdot u=1$, $S\cdot u=0$, $S\cdot S=-1$, and
orthogonality of the transported triad, and Table~\ref{tab:Sinv} collects the
measured bounds. One qualification belongs with them. They are evaluated at the
end of the pulse, and nothing is recorded along the trajectory, so an excursion
in the interior that came back would not appear.

\begin{table}[!tbp]
  \caption{Measured bounds on the exact invariants at the end of the pulse,
    taken over all configurations. The first two columns come from
    \texttt{code/spin\_magnitude.json}; the last is the same quantity over the
    wider grid behind FIG.~1 of the main text
    (\texttt{code/holonomy\_scan.json}), which reaches larger $\Anet$ and is
    correspondingly coarser. All entries measured.}
  \label{tab:Sinv}
  \squeezetable
  \footnotesize
  \begin{ruledtabular}
  \begin{tabular}{lccc}
    Quantity & $\gamma=1$ & $\gamma=5000$ & FIG.~1 grid \\
    \hline
    $|S\cdot u|$ & $2.24\times10^{-12}$ & $4.5\times10^{-7}$
      & $2.87\times10^{-11}$ \\
    $|S\cdot S+1|$ & $3.11\times10^{-12}$ & $9.0\times10^{-7}$
      & $5.61\times10^{-11}$ \\
    $\|R^{\!\top}R-\mathbb{1}\|$ & $1.65\times10^{-12}$
      & $9.0\times10^{-7}$ & --- \\
    $|u^{\mu}(+\infty)-u^{\mu}(-\infty)|$ & $1.24\times10^{-13}$
      & $1.24\times10^{-13}$ & --- \\
  \end{tabular}
  \end{ruledtabular}
\end{table}

Medians lie far below those maxima, $5.6\times10^{-15}$ for $|S\cdot u|$ and
$4.4\times10^{-15}$ for $|S\cdot S+1|$ at $\gamma=1$. Table~\ref{tab:Sinv}
reports worst cases over the whole scan. Truncation of the Gaussian envelope
enters at the same level, $|\bm{a}_{\perp}(\pm\eta_{\max})|\le
1.24\times10^{-13}$, and vanishes identically for $\cos^{2}$.

\begin{figure}[!tbp]
  \figorbox{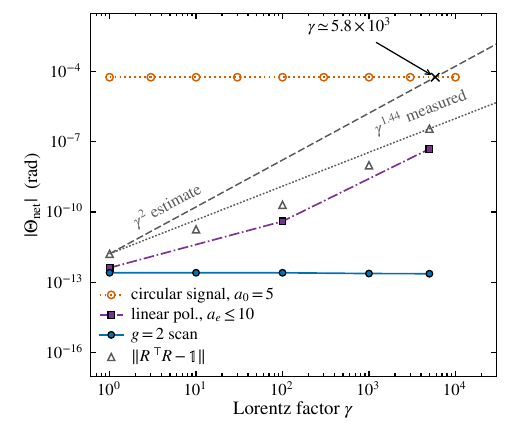}{FIG.~S1 artwork: measured net rotation against
    $\gamma$ for the two null families and for the circular-polarization
    signal, with $\gamma^{2}$ and $\gamma^{1.44}$ guide lines.}
  \caption{Measured net spin rotation against the Lorentz factor. Filled
    circles are the $g=2$ scan and filled squares the linear-polarization scan
    at inflated anomaly, both of which are zero by Eq.~(\ref{eq:Slambda0}) and
    by the fixed-axis argument of Sec.~\ref{sec:S2.2}, so what is plotted for
    them is round-off. The $g=2$ maximum is flat at
    $2.3$--$2.6\times10^{-13}$~rad, set by the extraction of the rotation
    vector from an $SO(3)$ matrix, and the growth with $\gamma$ is carried by
    the anomaly-inflated scan and by the measured invariant error
    $\|R^{\!\top}\!R-\mathbb{1}\|$ of the same runs (grey triangles), which
    rises from $1.65\times10^{-12}$ at $\gamma=1$ to $3.6\times10^{-7}$ at
    $\gamma=5\times10^{3}$. The dashed line is the $\gamma^{2}$ cancellation
    estimate anchored on the measured value at $\gamma=1$, and the dotted line
    is the power law that error actually follows; the data fall below the
    estimate by up to two decades, the effective exponent being $1.44$. Open
    symbols
    are the circularly polarized case at $a_{0}=5$, $N=1$, flat to six digits
    over four decades in $\gamma$. That estimate meets the signal at the cross
    near $\gamma\simeq5.8\times10^{3}$, which marks where the laboratory
    formulation stops resolving it.}
  \label{fig:Sfloor}
\end{figure}

The invariant error grows with energy, because the laboratory components of $S$
and $u$ scale as $\gamma$ while the invariants are built from cancellations
among them. A cancellation estimate would give $\gamma^{2}\epsilon_{\rm mach}$,
and the measured growth is slower than that: from $1.65\times10^{-12}$ at
$\gamma=1$ to $3.6\times10^{-7}$ at $\gamma=5\times10^{3}$, an effective
exponent of $1.44$ over the range, approaching $2$ only above
$\gamma\simeq10^{2}$. The net rotation at $g=2$ behaves differently, and the
difference matters for how the null results are read, since that quantity does
not grow with $\gamma$ at all but stays flat at
$2.3$--$2.6\times10^{-13}$~rad across the whole range, being set by the
extraction of a rotation vector from a matrix close to the identity. It scales
with intensity instead, the median rising roughly as $a_{0}$, from
$7.4\times10^{-16}$ at $a_{0}=0.1$ through $5.0\times10^{-15}$ at $a_{0}=1$ to
$5.0\times10^{-14}$ at $a_{0}=10$, which is the signature of accumulated
round-off. Figure~\ref{fig:Sfloor} separates the two behaviors. The precision
scans were run at $\gamma=1$, which Sec.~\ref{sec:S3.2} shows to be
equivalent.

\subsection{Convergence in the integrator tolerance}
\label{sec:S4.3}

A null result is worth only as much as the evidence that the step size has
stopped mattering, so the plane-wave zero was rerun over the integrator
tolerance at $\ano=0$, with ${\rm rtol}$ from $10^{-9}$ to $10^{-14}$ and
${\rm atol}=10^{-3}{\rm rtol}$. Table~\ref{tab:Stol} gives three
configurations, and they read in opposite directions.

\begin{table*}[!tbp]
  \caption{Measured $|\Theta_{\rm net}|$ in radians at $g=2$ against the
    relative tolerance of the DOP853 integrator, at
    ${\rm atol}=10^{-3}{\rm rtol}$. The working tolerance of the runs reported
    elsewhere is ${\rm rtol}=10^{-13}$.
    Each row is at one carrier-envelope phase, $\varphi_{0}=\pi/2$ for the
    first and $\varphi_{0}=0.7$ for the other two. SciPy raises any relative
    tolerance below $100$ machine epsilons, so the last column is executed at
    $2.22\times10^{-14}$. Sixteen of the eighteen entries reproduce an earlier
    run to the digits printed. The two that do not are both in the loosest
    column, $10^{-9}$: the third row differs by rounding, and the second
    returns $9.8\times10^{-14}$ against $3.0\times10^{-13}$ before, a factor of
    three whose original configuration could not be recovered. Both cells
    carry the recomputed value, and neither enters any bound quoted here.
    Source: \texttt{code/spin\_magnitude.json}, key \texttt{t6}.}
  \label{tab:Stol}
  \squeezetable
  \footnotesize
  \begin{ruledtabular}
  \begin{tabular}{lcccccc}
    Case & $10^{-9}$ & $10^{-10}$ & $10^{-11}$ & $10^{-12}$ & $10^{-13}$
      & $10^{-14}$ \\
    \hline
    $a_{0}=10$, $N=1$, Gauss, $\gamma=10$
      & $3.4\times10^{-11}$ & $6.0\times10^{-12}$ & $5.5\times10^{-13}$
      & $2.0\times10^{-13}$ & $2.6\times10^{-13}$ & $2.4\times10^{-13}$ \\
    $a_{0}=10$, $N=4$, Gauss, $\gamma=1$
      & $9.8\times10^{-14}$ & $3.6\times10^{-13}$ & $7.9\times10^{-14}$
      & $1.0\times10^{-13}$ & $2.6\times10^{-13}$ & $1.2\times10^{-13}$ \\
    $a_{0}=1$, $N=1$, $\cos^{2}$, $\gamma=1$
      & $4.4\times10^{-14}$ & $6.6\times10^{-15}$ & $3.3\times10^{-15}$
      & $1.7\times10^{-15}$ & $5.0\times10^{-16}$ & $2.3\times10^{-16}$ \\
  \end{tabular}
  \end{ruledtabular}
\end{table*}

At $a_{0}=10$ the zero has reached the arithmetic. Tightening the tolerance by
two decades from $10^{-12}$ to $10^{-14}$ does not lower $\Theta_{\rm net}$,
which instead fluctuates over $(1.0$--$2.6)\times10^{-13}$ while the number of
right-hand-side evaluations triples, and that is the evidence that the
vanishing is a property of the equations. The same fact costs us two digits of
the headline number. At ${\rm rtol}=10^{-12}$ the identical computation returns
$2.0\times10^{-13}$ and at $10^{-14}$ it returns $2.4\times10^{-13}$, a spread
of $22.5\%$ below the value $2.57\times10^{-13}$ obtained at the working
tolerance, so three significant figures are not secured and the defensible
statement is the bound $\le3\times10^{-13}$~rad used in
Table~\ref{tab:Svalid}. At $a_{0}\lesssim1$ the picture inverts: the measured
value still falls monotonically with the tolerance, from $4.4\times10^{-14}$ to
$2.3\times10^{-16}$, so over that part of the grid the reported zero is an
upper bound set by truncation and not a round-off floor. Little room is left in
either direction, since SciPy clamps ${\rm rtol}$ below
$2.22\times10^{-14}$ and the working value already sits one decade above that.

\subsection{Validation runs}

Table~\ref{tab:Svalid} collects the tests. The entry that carries the most
weight is the second: the vanishing at $g=2$ survives when the anomaly is
switched on and inflated by four orders of magnitude above its physical value,
which rules out the reading ``the effect is there but small.''

\begin{table*}[!tbp]
  \caption{Validation of the plane-wave results. All bounds are measured.
    Cases counted are independent parameter combinations; the scan covers two
    envelopes, $a_{0}$ from $0.1$ to $10$, $N$ from $0.5$ to $16$, Lorentz
    factors from $1$ to $5\times10^{3}$, three carrier-envelope phases and
    ellipticities from $0$ to $1$. The zero-area row is measured on the 18
    such configurations of the grid behind FIG.~1
    (\texttt{code/holonomy\_scan.json}); the value in parentheses there is the
    covariant cross-check of Sec.~\ref{sec:S4} on four cases of its own, at
    the coarser floor described in that section. Source data:
    \texttt{code/spin\_magnitude.json} (keys \texttt{t1}, \texttt{t3}),
    \texttt{code/holonomy\_scan.json} and
    \texttt{code/indep\_check*.py}.}
  \label{tab:Svalid}
  \squeezetable
  \footnotesize
  \begin{ruledtabular}
  \begin{tabular}{lcl}
    Test & Cases & Measured bound \\
    \hline
    Net rotation at $g=2$
      & 180 & $\max|\Theta_{\rm net}|\le3\times10^{-13}$~rad
             ($2.57\times10^{-13}$ at ${\rm rtol}=10^{-13}$) \\
    Net rotation, linear polarization, anomaly up to $\ano=10$
      & 60 & $4.90\times10^{-8}$~rad at $\gamma=5000$,
             $\le4.1\times10^{-11}$~rad at $\gamma\le10^{2}$ \\
    Closed-form solution against the integrator, three interior phases
      & 18 & max relative error $7.88\times10^{-13}$ \\
    Interaction-picture generator against Eq.~(\ref{eq:SM}) on a phase grid
      & --- & $\max|M-\nhat\wedge a'|=2.7\times10^{-15}$ \\
    Independence of $\gamma$, circular polarization, $\gamma=1$--$10^{4}$
      & 9 & 6 matching significant digits \\
    Reparameterization of the phase at fixed curve, $\ano$ up to $0.25$
      & 8 & 8--9 matching significant digits \\
    Zero-area curve $a_{y}\propto a_{x}^{2}$, $\ano$ up to $0.6$
      & 18 & $\le9.95\times10^{-12}$~rad
             ($\le7.7\times10^{-11}$ on the covariant check) \\
    Return of the four-velocity
      & all & $|u(+\infty)-u(-\infty)|\le1.24\times10^{-13}$ \\
  \end{tabular}
  \end{ruledtabular}
\end{table*}

\subsection{Scaling laws}
\label{sec:S4.4}

How well Eq.~(\ref{eq:Sresult}) reproduces the measured angle away from linear
polarization depends on which scan is meant, and the two answers differ by two
digits. Over the wide grid behind FIG.~1 of the main text, which mixes
envelopes,
ellipticities, chirps and polarization gating, the relative residual has median
$8.0\times10^{-5}$ and maximum $9.1\times10^{-4}$, that is four digits
typically and three at worst. Along a single parameter axis the agreement is
the five digits quoted in the main text, and Table~\ref{tab:Sscaling} gives
the
ratio of measurement to prediction axis by axis. The ratio
$\Theta_{\rm net}/\ano^{2}$ stays constant to $0.8\%$ across a $26$-fold range
in
the anomaly, which is the direct statement that the effect is quadratic. Points
with $\ano\ge0.1$ are excluded there. Beyond that value $\Theta_{\rm net}$
exceeds $\pi$ and the extracted rotation vector wraps, in the sense of
Sec.~\ref{sec:S2.3}.

\begin{table}[!tbp]
  \caption{Measured net rotation divided by $-\tfrac12\ano^{2}\Anet$, along one
    parameter axis at a time. Circular polarization, Gaussian envelope,
    $\gamma=10^{3}$; the fixed values are $a_{0}=10$ and $N=4$, except in the
    first row, where $N=1$. The last row
    instead lists $\Theta_{\rm net}/\ano^{2}$, which Eq.~(\ref{eq:Sresult})
    requires to equal $\Anet/2=668.8$. Source:
    \texttt{code/spin\_magnitude.json}, key \texttt{t3}.}
  \label{tab:Sscaling}
  \squeezetable
  \footnotesize
  \begin{ruledtabular}
  \begin{tabular}{lcc}
    Axis & Range & Ratio \\
    \hline
    $a_{0}$ & $0.1$--$50$ & $1.00000$--$0.99971$ \\
    $N$ & $0.5$--$16$ & $0.99999$ \\
    $\delta$ & $0.05$--$1$ & 5 digits \\
    $\delta=0$ & --- & $8.7\times10^{-12}$~rad \\
    $\ano$, as $\Theta/\ano^{2}$ & $1.2\times10^{-3}$--$3\times10^{-2}$
      & $668.8$--$663.6$ \\
  \end{tabular}
  \end{ruledtabular}
\end{table}

For a circularly polarized Gaussian pulse the law evaluates to
$\Theta_{\rm net}=2.2486\times10^{-6}a_{0}^{2}N$~rad, and for $\cos^{2}$ to
$2.1759\times10^{-6}a_{0}^{2}N$~rad. The $3.2\%$ difference is entirely
$\int f^{2}d\eta$: at $a_{0}=10$ and $N=1$ the two areas are
$\Anet=334.412$ and $323.603$ (measured).

\section{Finite focusing}
\label{sec:S5}

\subsection{Field model}
\label{sec:S5.1}

Equation~(\ref{eq:Sresult}) rests on $A=A(\eta)$, which no focused beam
satisfies. The field used below is a Gaussian pulse built from the
Lax--Louisell--McKnight (LLM) expansion in the diffraction parameter
$\varepsilon=1/(kw_{0})=\lambda/2\pi w_{0}$, in the form standard in the
strong-field literature~\cite{Lax1975,Salamin2002,Salamin2007}. With
$s=\varepsilon^{2}(X^{2}+Y^{2})$, $\zeta=2\varepsilon^{2}Z$ and
$F=1/(1+i\zeta)$, the transverse complex amplitude is carried to third order,
\begin{equation}
  \psi=\psi_{0}+\varepsilon^{2}\psi_{2},
  \quad
  \psi_{0}=F e^{-Fs},
  \label{eq:Spsi}
\end{equation}
\begin{equation}
  \psi_{2}=e^{-Fs}\bigl[c_{\psi}F^{2}+(2-c_{\psi})F^{3}s-F^{4}s^{2}\bigr].
  \label{eq:Spsi2}
\end{equation}
We rederived $\psi_{2}$ rather than copying it: substitution into the wave
equation gives
$s\psi_{2,ss}+\psi_{2,s}+i\psi_{2,\zeta}=-\partial_{\zeta}^{2}\psi_{0}$,
whose general solution is Eq.~(\ref{eq:Spsi2}) with one free constant
$c_{\psi}$ multiplying $\partial_{\zeta}\psi_{0}$, a homogeneous paraxial
solution. That constant is the $O(\varepsilon^{2})$ waist-shift ambiguity
built into the expansion. We work at $c_{\psi}=0$, where the correction
vanishes on axis and $a_{0}$ keeps its meaning as the peak potential at focus.
The spread over $c_{\psi}$ is carried as a systematic error.

The longitudinal component is obtained by closing the Coulomb gauge
$\nabla\!\cdot\!\bm{a}=0$ exactly, by the fixed-point iteration
$W=i(Dg+\partial_{Z}W)$ with $W_{0}=iDg$ and
$D=\alpha_{x}\psi_{,X}+\alpha_{y}\psi_{,Y}$. Two iterations give a
longitudinal field complete through $\varepsilon^{3}$ and through
$\varepsilon/\sigma^{2}$. Then $\bm{E}=-\partial_{t}\bm{a}$ and
$\bm{B}=\nabla\times\bm{a}$. At retained order 1 the measured ratio
$E_{z}/E_{\perp}$ grows as $\varepsilon^{1.07}$ from $0.0174$ at
$\varepsilon=0.025$, and with $\psi_{2}$ retained it reaches $0.276$ at
$\varepsilon=0.30$, so the model carries a longitudinal field of the right
order.

\subsection{Maxwell residuals}
\label{sec:S5.2}

Two of Maxwell's equations hold identically, because the fields are built from
a potential: $\nabla\!\cdot\!\bm{B}$ and
$\nabla\times\bm{E}+\partial_{t}\bm{B}$ are machine zero at all sampled
points. The other two are measured on 48 points fixed in the self-similar
variables $(\rho/w_{0},z/z_{R},\eta/\sigma)$, so that different $\varepsilon$
are compared at the same place in the beam, with residuals normalized to
$a_{0}$. Table~\ref{tab:Smaxwell} gives them for a quasimonochromatic run,
$\sigma=100$, which isolates the LLM hierarchy from the envelope.

\begin{table}[!tbp]
  \caption{Maxwell residuals of the field model, normalized to $a_{0}$, for a
    quasimonochromatic pulse ($\sigma=100$) at retained order 1 and order 3 in
    $\varepsilon$. $\nabla\!\cdot\!\bm{B}=0$ and Faraday's law are satisfied
    identically and are not listed. All entries measured. Source:
    \texttt{code/focused\_beam.json}, stage \texttt{field}.}
  \label{tab:Smaxwell}
  \squeezetable
  \footnotesize
  \begin{ruledtabular}
  \begin{tabular}{ccccc}
    & \multicolumn{2}{c}{$\nabla\!\cdot\!\bm{E}$}
    & \multicolumn{2}{c}{$\nabla\times\bm{B}-\partial_{t}\bm{E}$}\\
    \cline{2-3}\cline{4-5}
    $\varepsilon$ & order 1 & order 3 & order 1 & order 3 \\
    \hline
    0.30 & $1.07\times10^{-2}$ & $1.05\times10^{-2}$
         & $3.87\times10^{-2}$ & $1.69\times10^{-2}$ \\
    0.20 & $6.30\times10^{-4}$ & $6.23\times10^{-4}$
         & $8.35\times10^{-3}$ & $7.35\times10^{-4}$ \\
    0.15 & $8.50\times10^{-5}$ & $8.44\times10^{-5}$
         & $2.79\times10^{-3}$ & $2.98\times10^{-4}$ \\
    0.10 & $5.29\times10^{-6}$ & $5.27\times10^{-6}$
         & $5.63\times10^{-4}$ & $1.21\times10^{-4}$ \\
    0.05 & $8.82\times10^{-8}$ & $8.80\times10^{-8}$
         & $4.75\times10^{-5}$ & $3.01\times10^{-5}$ \\
  \end{tabular}
  \end{ruledtabular}
\end{table}

The Gauss residual falls as $\varepsilon^{5.5}$, a consequence of the exact
Coulomb closure, and the Amp\`ere residual as $\varepsilon^{3.5}$ at order 1.
Keeping $\psi_{2}$ lowers the Amp\`ere residual by a factor $4.6$ to $11$ over
$\varepsilon=0.1$--$0.2$, which is the hierarchy working as intended.

For a physical pulse the picture is worse and we state it plainly. At
$\sigma=3.77$ ($N=1$) the residuals fall only as $\varepsilon^{1.4}$ for Gauss
and $\varepsilon^{2}$ for Amp\`ere, and at $\varepsilon=0.15$ they reach
$1.4\times10^{-3}$ and $7.2\times10^{-3}$ of $a_{0}$. The finite spectral width
produces terms $\propto\varepsilon^{p}/\sigma^{q}$ from the product of the
$z$-dependent mode with the envelope in $\eta$, and no finite order in
$\varepsilon$ removes them, so this is a property of the pulse and not a defect
of the LLM hierarchy. An isodiffracting envelope would remove them, and we did
not try one. The practical consequence appears in Sec.~\ref{sec:S5.6}: any
measured quantity below $10^{-3}a_{0}$ is a property of the model.

In the opposite limit the model reduces correctly. At $\varepsilon\le10^{-2}$
the field agrees with the exact plane wave of Eq.~(\ref{eq:Senvelope}) to
$1.1\times10^{-16}$ at five phases (measured).

\subsection{Dynamics and observable}

The Lorentz and BMT equations are integrated together in laboratory time,
\begin{align}
  \dot{\bm{x}}&=\bm{\beta},
  \qquad
  \dot{\bm{u}}=\bm{E}+\bm{\beta}\times\bm{B},\nonumber\\
  \dot{S}^{\mu}&=\frac1\gamma
    \Bigl[(1+\ano)(fS)^{\mu}+\ano\,u^{\mu}(S\!\cdot\!fu)\Bigr],
  \label{eq:Sfocdyn}
\end{align}
with $u^{0}=\sqrt{1+\bm{u}^{2}}$ imposed algebraically, so $u\cdot u=1$ holds
exactly. Working tolerances are ${\rm rtol}=10^{-11}$ and
${\rm atol}=10^{-13}$, with ${\rm rtol}=10^{-13}$ as the convergence
reference. Along a trajectory sampled at 401 points we measured
$|u\cdot u-1|\le2.8\times10^{-14}$, $|S\cdot u|\le2.8\times10^{-12}$ and
$|S\cdot S+1|\le2.3\times10^{-12}$, six to seven orders below every quantity
reported here.

The observable is therefore the spin rotation relative to the momentum. Three
rest-frame four-vectors are transported, a reference triad is constructed at
the exit by boosting the entrance triad along $u_{0}\to u_{f}$, and the
rotation vector $r$ is read from
$R_{ij}=-\langle\tilde{e}_{i},e_{j}(t_{f})\rangle$. Removing that boost removes
with it the kinematic transport that accompanies ponderomotive deflection,
which would otherwise dominate everything. In the plane-wave limit
$u_{f}=u_{0}$, the boost is the identity, and $r$ reduces to the net rotation
of Sec.~\ref{sec:S4}.

Five triads sharing a single orbit, at anomalies $\ano\in\{0,\pm h,\pm2h\}$ with
$h=0.05$, are separated by five-point differences into
\begin{equation}
  r(\ano)=r_{0}+\ano\,r_{1}+\ano^{2}r_{2}+O(\ano^{3}) .
  \label{eq:Sstratify}
\end{equation}
Both $r_{0}$ and $r_{1}$ vanish identically in a plane wave, the first by
Eq.~(\ref{eq:Slambda0}) and the second by the argument of
Sec.~\ref{sec:S2.2}.

\subsection{The limit $\varepsilon\to0$}
\label{sec:S5.4}

Table~\ref{tab:Seps} gives the scan. All three violations enter at order
$\varepsilon^{2}$, and the range over which that is true has to be stated with
the exponents. Over $\varepsilon=0.025$--$0.15$ the local log-log slopes run
$2.71\to2.02$ for $|r_{0}|$, $2.71\to2.14$ for $|r_{1}|$, $1.92\to2.00$ for
the deviation of $r_{2}$ from its plane-wave value, and $3.11\to2.06$ for the
transverse momentum kick $\Delta u_{\perp}$. Over the whole scanned range the
fitted exponents at $\gamma=10$ are $2.87$ for $|r_{0}|$ and $2.89$ for
$|r_{1}|$, and the local slopes on the last two intervals reach $4.0$ and
$6.3$, so the $\varepsilon^{2}$ law is a statement about
$\varepsilon\lesssim0.15$ and the two widest points of Table~\ref{tab:Seps}
are order-of-magnitude entries. The deviation of the holonomic coefficient is
\begin{equation}
  \frac{|r_{2}+\tfrac12\Anet\nhat|}{\tfrac12\Anet}\simeq1.05\,\varepsilon^{2}
  \qquad(\gamma=10),
  \label{eq:Sr2dev}
\end{equation}
measured as $1.06,\,1.05,\,1.05,\,1.04,\,1.03,\,0.995$ at
$\varepsilon=0.025$--$0.15$. At $\gamma=1$ the same law holds with
coefficient $3.4$. The coefficient, unlike the exponent, carries a systematic
error of a factor of about three from the field model itself, and
Sec.~\ref{sec:S5.7} propagates it.

\begin{table}[!tbp]
  \caption{Focused Gaussian pulse: circular polarization, $a_{0}=1$, $N=1$,
    $\varphi_{0}=0$, head-on collision at $\gamma=10$. The plane-wave
    reference is $-\tfrac12\Anet=-1.6720609$. All entries measured. Source:
    \texttt{code/focused\_beam.json}, stage \texttt{eps}.}
  \label{tab:Seps}
  \squeezetable
  \footnotesize
  \begin{ruledtabular}
  \begin{tabular}{cccccc}
    $\varepsilon$ & $w_{0}/\lambda$ & $|r_{0}|$ & $|r_{1}|$ & $r_{2,z}$
      & dev.\ of $r_{2}$ \\
    \hline
    0.300 & 0.53 & $3.24\times10^{-2}$ & $3.22\times10^{-2}$
          & $-1.79960$ & $7.82\times10^{-2}$ \\
    0.200 & 0.80 & $2.55\times10^{-3}$ & $2.53\times10^{-3}$
          & $-1.73559$ & $3.80\times10^{-2}$ \\
    0.150 & 1.06 & $8.04\times10^{-4}$ & $8.05\times10^{-4}$
          & $-1.70950$ & $2.24\times10^{-2}$ \\
    0.100 & 1.59 & $2.67\times10^{-4}$ & $2.68\times10^{-4}$
          & $-1.68928$ & $1.03\times10^{-2}$ \\
    0.070 & 2.27 & $1.20\times10^{-4}$ & $1.20\times10^{-4}$
          & $-1.68062$ & $5.12\times10^{-3}$ \\
    0.050 & 3.18 & $5.91\times10^{-5}$ & $5.83\times10^{-5}$
          & $-1.67646$ & $2.63\times10^{-3}$ \\
    0.035 & 4.55 & $2.85\times10^{-5}$ & $2.75\times10^{-5}$
          & $-1.67422$ & $1.29\times10^{-3}$ \\
    0.025 & 6.37 & $1.44\times10^{-5}$ & $1.34\times10^{-5}$
          & $-1.67316$ & $6.60\times10^{-4}$ \\
  \end{tabular}
  \end{ruledtabular}
\end{table}

For linear polarization $r_{2}$ stays at $\le7.5\times10^{-11}$ for every
$\varepsilon$ up to $0.30$ (measured). Nothing in Eq.~(\ref{eq:Sresult})
guarantees that outside the plane-wave limit, and we have no argument for why
it holds. It is an observation.

A direct run at the physical anomaly $\ano=1.15965\times10^{-3}$, without finite
differences, gives the full anomalous observable. The ratio to the plane-wave
value is $1.0067$, $1.0271$ and $1.3033$ at $\varepsilon=0.025$, $0.050$ and
$0.150$, that is
\begin{equation}
  \frac{\Theta_{\rm anom}(\varepsilon)}{\tfrac12\ano^{2}\Anet}-1
  \simeq\frac{10.7}{a_{0}}\,\varepsilon^{2}
  \qquad(\gamma=10),
  \label{eq:Sepsviolation}
\end{equation}
with the coefficient measured as $10.7$, $10.8$ and $13.5$. Equation~(13) of
the main text
quotes the same numbers. The deviation is accounted for by the recovered term
$\ano\,r_{1}$ alone, since
$c_{1}\varepsilon^{2}/(\ano\Anet/2)=11.0\,\varepsilon^{2}$
from the measured $c_{1}$.

The amplitudes depend strongly on energy while the exponents do not. Measured
$|r_{0}|=c_{0}\varepsilon^{2}$ and $|r_{1}|=c_{1}\varepsilon^{2}$ give
$c_{0}=7.22$, $c_{1}=7.48$ at $\gamma=1$; $c_{0}=0.0230$, $c_{1}=0.0214$ at
$\gamma=10$; and $c_{0}=0.0165$, $c_{1}=0.0131$ at $\gamma=10^{3}$, a
436-fold range in $c_{0}$. Translating Eq.~(\ref{eq:Sepsviolation}) into a
waist requirement gives Table~\ref{tab:Swindow}.

\begin{table}[!tbp]
  \caption{Waist above which Eq.~(\ref{eq:Sresult}) holds to the stated
    accuracy, at $a_{0}\simeq1$, $N=1$, circular polarization, from
    $\varepsilon<\sqrt{{\rm acc.}/c}$ and
    $w_{0}/\lambda=1/(2\pi\varepsilon)$. The $\gamma=10$ column uses the
    directly measured $c=10.7$ of Eq.~(\ref{eq:Sepsviolation}); the other two
    are estimated from the measured $c_{1}$, giving $c=3.86\times10^{3}$ and
    $c=6.8$. Every entry is a lower bound for the field model at
    $c_{\psi}=0$, and Sec.~\ref{sec:S5.7} shows that $c$ carries a systematic
    factor of about three, which multiplies each waist by $\sqrt{3}$: the
    $1\%$ row then reads $171\lambda$, $9.0\lambda$ and $7.2\lambda$.}
  \label{tab:Swindow}
  \begin{ruledtabular}
  \begin{tabular}{lccc}
    Accuracy & $\gamma=1$ & $\gamma=10$ & $\gamma=10^{3}$ \\
    \hline
    $10\%$ & $w_{0}>31\lambda$ & $w_{0}>1.6\lambda$ & $w_{0}>1.3\lambda$ \\
    $1\%$  & $w_{0}>99\lambda$ & $w_{0}>5.2\lambda$ & $w_{0}>4.1\lambda$ \\
  \end{tabular}
  \end{ruledtabular}
\end{table}

The dependence on intensity runs the other way. The measured ratio
$|\ano\,r_{1}|/|\ano^{2}r_{2}|$ at $\varepsilon=0.15$ and $\gamma=10$ is
$1.56,\,0.79,\,0.41,\,0.25,\,0.28$ at $a_{0}=0.25,\,0.5,\,1,\,2,\,4$, falling
as $1/a_{0}$ up to $a_{0}\simeq2$ and flattening beyond it. Therefore the
holonomy formula gains accuracy as the intensity rises, by $1/a_{0}$ over
the range where the law holds, and we have not scanned $\varepsilon$ at
$a_{0}\gtrsim10$.

\subsection{What the dominant term is}
\label{sec:S5.5}

The largest of the three coefficients is $r_{0}$, and it carries no
information about $g$, since it is computed from the orbit alone. At
$\gamma=10^{3}$ we measured $|r_{0}|=|\Delta u_{\perp}|$ to four digits with an
axis orthogonal to the momentum, $|r_{0\parallel}|/|r_{0\perp}|\le
6\times10^{-5}$, and that identifies $r_{0}$ as the Thomas--Wigner
rotation~\cite{Thomas1926,Wigner1939} accompanying the ponderomotive kick,
whose existence in a focused beam follows from the failure of the
Lawson--Woodward theorem~\cite{Lawson1979,Esarey1995}. The separation of the
axes is already much weaker one decade down, $5.7\times10^{-3}$ at
$\gamma=10^{2}$, and it fails outright below $\gamma\simeq3$, where $r_{0}$
becomes mostly parallel to the momentum ($0.148$ against $0.014$ transverse at
$\gamma=1$, $\varepsilon=0.15$, $a_{0}=1$) and exceeds $\Delta u_{\perp}$ by a
factor $10$ to $50$ because the electric term of Eq.~(\ref{eq:Slab})
dominates. That is the same corner of the parameter space in which
Sec.~\ref{sec:S6} places the best classical working point, and the two
requirements pull against each other.

The net rotation here is nonzero already at $g=2$, and the axis lies
perpendicular to the momentum, so the projection of the spin on the momentum
loses the protection it had in a plane wave. A
longitudinally polarized beam does depolarize in a focused pulse, by
$8.0\times10^{-4}$~rad at $a_{0}=1$, $\varepsilon=0.15$, $\gamma=10$ and by up
to $0.27$~rad at $a_{0}=16$ (measured). We have not searched the literature on
$g=2$ spin rotation in focused beams, and this identification should be read
as a measurement of ours rather than as a claim of novelty.

We measured $|r_{1}|/|r_{0}|$ between $0.79$ and $1.07$ over the
$\varepsilon\le0.15$ part of the grid, the ratio reaching $1.21$ at
$\gamma=1$ and $\varepsilon=0.30$. The two coefficients are produced by the
same mechanism. The focused geometry offers no selectivity for the anomaly.

\subsection{Recovered carrier-envelope-phase dependence}
\label{sec:S5.6}

For circular polarization and an electron on axis the CEP dependence is absent
in the focused beam as well: the relative modulation over 16 values of
$\varphi_{0}$ is $4.3\times10^{-12}$ for $r_{0}$, $3.2\times10^{-10}$ for
$r_{1}$ and $2.3\times10^{-12}$ for $r_{2}$ (measured). The reason is exact and
identical to Sec.~\ref{sec:S3.6}: for
$\alpha=(1,i)/\sqrt2$ and a cylindrically symmetric $\psi$, rotating the whole
configuration about the beam axis is the same as shifting $\varphi_{0}$. Off
axis the symmetry breaks, and the relative modulation of $|r_{0}|$ reaches
$22\%$ for circular polarization and $35\%$ for linear polarization at an
impact parameter $b=1.5w_{0}$.

\begin{figure}[!tbp]
  \figorbox{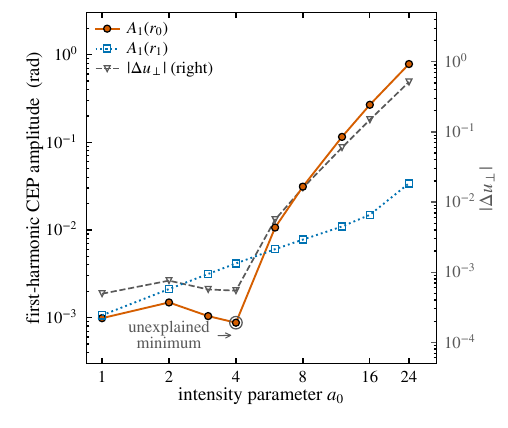}{FIG.~S2 artwork: CEP amplitude of the focused-beam spin
    rotation against $a_{0}$, for the zeroth- and first-order coefficients,
    with the ponderomotive kick on a second axis.}
  \caption{Carrier-envelope-phase amplitude of the spin rotation relative to
    the momentum, for linear polarization at $\varepsilon=0.15$
    ($w_{0}=1.06\lambda$), $\gamma=10$, $N=1$, electron on axis. Filled
    symbols: first-harmonic amplitude of the $g$-independent coefficient
    $r_{0}$. Open symbols: the same for the first-order coefficient $r_{1}$,
    which is linear in $a_{0}$ up to $a_{0}\simeq12$ (measured local slopes
    $0.99$ down to $0.85$) and steepens to $2.0$ over the last interval. Open
    downward triangles on the right-hand axis are the transverse momentum kick
    $|\Delta u_{\perp}|$ of the same runs, showing how the ponderomotive
    ejection grows alongside the signal. The dependence is a pure first
    harmonic, with the second harmonic below $2\times10^{-12}$ of the first
    for $r_{0}$. The annotated point near $a_{0}=4$ is a minimum of
    $A_{1}(r_{0})$ that we cannot explain. At $a_{0}=24$ the rotation swings by
    $1.56$~rad between $\varphi_{0}$ and $\varphi_{0}+\pi$, while the
    anomalous part of the same signal is $3.9\times10^{-5}$~rad.}
  \label{fig:Scep}
\end{figure}

For linear polarization the CEP controls the sign and magnitude of the whole
rotation. The vector $r$ lies along the normal to the polarization plane, the
mean over $\varphi_{0}$ is zero, and the dependence is a pure first harmonic,
with $A_{2}/A_{1}\le2\times10^{-12}$ for $r_{0}$. The bound is
$3.7\times10^{-12}$ for $r_{1}$ and only $1.1\times10^{-2}$ for $r_{2}$, whose
first harmonic is small enough that the second is not negligible beside it. The
measured amplitude is
$A_{1}(r_{0})\simeq0.0209\,\varepsilon^{2}$~rad and
$A_{1}(r_{1})\simeq0.0235\,\varepsilon^{2}$~rad at $a_{0}=1$, $\gamma=10$,
$N=1$, with local slopes $2.46\to2.05$. Figure~\ref{fig:Scep} shows the growth
with intensity: $A_{1}(r_{0})$ is linear in $a_{0}$ up to $a_{0}\simeq2$,
passes through a minimum near $a_{0}\simeq4$ that we cannot explain, and then
grows as $a_{0}^{2.6}$--$a_{0}^{3.2}$, reaching $0.781$~rad at $a_{0}=24$ and
$\varepsilon=0.15$. The first-order amplitude $A_{1}(r_{1})$ is linear in
$a_{0}$ only up to $a_{0}\simeq12$: the measured local slopes fall from
$0.99$ to $0.85$ over $a_{0}=1$--$12$ and then rise to $1.06$ and $2.04$ over
the last two intervals.

The anomalous part of the signal at the record point $a_{0}=24$ is
$3.9\times10^{-5}$~rad, that is $5\times10^{-5}$ of the total. At $a_{0}=1$,
where the rest of this section works, the same fraction is $1.3\times10^{-3}$,
twenty-five times larger. Either way that part shares the CEP harmonic, the
polarization signature and the $\gamma$ dependence of the $g$-independent one,
so it cannot be separated from it. The amplitude also collapses with pulse
length: measured $A_{1}(r_{0})$ at $\varepsilon=0.15$, $a_{0}=1$, $\gamma=10$
is $6.29\times10^{-2}$, $9.84\times10^{-4}$, $2.63\times10^{-7}$ and
$1.36\times10^{-11}$~rad at $N=0.5,1,2,4$, an exponential collapse whose local
rate is not constant, steepening to about $e^{-2.2\sigma}$ in the middle and
flattening to $e^{-1.3\sigma}$ over the last interval. The last two values lie
below the Maxwell residuals of Sec.~\ref{sec:S5.2} and should be read as zero
within the model.

\subsection{Systematic error of the field model}
\label{sec:S5.7}

Three variants of the field were run through identical dynamics: order 3 with
$c_{\psi}=0$, order 3 with $c_{\psi}=1$, and order 1. The spread in $|r_{0}|$
is $18\%$ at $\varepsilon=0.20$, $3.2\%$ at $\varepsilon=0.10$ and $0.8\%$ at
$\varepsilon=0.05$ (measured at $\gamma=10$, $a_{0}=1$, $N=1$,
$\varphi_{0}=0.3$). That spread itself scales as $\varepsilon^{2}$, so it
enters the $\varepsilon^{4}$ correction and leaves the exponent alone, while
the amplitudes $c_{0}$ and $c_{1}$ carry it in full. Numbers at
$\varepsilon\ge0.2$ are order-of-magnitude statements. The quantitative range
is $\varepsilon\le0.15$.

On the deviation of $r_{2}$, the quantity that Eq.~(\ref{eq:Sr2dev}) and
Eq.~(\ref{eq:Sepsviolation}) are built from, the systematic behaves quite
differently and worse. All three variants give an exponent of exactly $2$, so
that part is firm, but the coefficient does not converge as
$\varepsilon\to0$. Measured values of the deviation divided by
$\varepsilon^{2}$ are $0.950$, $1.030$ and $1.051$ at
$\varepsilon=0.20$, $0.10$ and $0.05$ for $c_{\psi}=0$; $3.130$, $3.072$ and
$3.062$ for $c_{\psi}=1$; and $1.158$, $1.074$ and $1.062$ at order 1. The
ratio between the extremes is $3.3$, $3.0$ and $2.9$, a factor of about three
that stays put as $\varepsilon$ shrinks. The coefficient $1.05$ of
Eq.~(\ref{eq:Sr2dev}) and the coefficient $10.7/a_{0}$ of
Eq.~(\ref{eq:Sepsviolation}) are therefore quoted for one convention of the
field model and are uncertain by that factor, which is why
Table~\ref{tab:Swindow} carries the $\sqrt{3}$ correction to every waist. We
work at $c_{\psi}=0$ because the correction then vanishes on axis and $a_{0}$
keeps its meaning as the peak potential at focus, and we know of no argument
that makes that choice physical.

\subsection{What was not computed}
\label{sec:S5.8}

Exact nonparaxial fields, such as the complex-source or Sewell--Barnett
solutions, were not used and the LLM model was never checked against
them~\cite{DiPiazza2021}, only the Gaussian envelope was run in the focused
geometry, and neither $\cos^{2}$ nor an isodiffracting envelope, which is the
one construction that would remove the $\varepsilon^{p}/\sigma^{q}$ terms of
Sec.~\ref{sec:S5.2}, was tried at all. Radiation reaction was not
integrated in the focused geometry any more than in the plane wave. Averaging
over the focal volume and over the beam spectrum was not done: every number
above is for a single electron at a fixed impact parameter with exact
synchronization, and the measured spread over $b$ within one waist is a factor
of $20$ in $|r_{0}|$. Ellipticity at $\varphi_{0}\neq0$, a transverse offset
along $y$, oblique geometries, chirp, polarization gating, and higher
transverse modes were not scanned.

\section{Regime of validity}
\label{sec:S6}

Everything below is estimated from published scalings. We integrated no
radiative process, classical or quantum.

The quantum nonlinearity parameter is
\begin{equation}
  \chi=\frac{\hbar\omega}{mc^{2}}\,\kappa\,a_{0},
  \qquad \kappa=k\cdot u ,
  \label{eq:Schi}
\end{equation}
with $k$ normalized to unit frequency, so that $\kappa\simeq2\gamma$ head-on
and $\chi\simeq6.07\times10^{-6}\gamma a_{0}$ at $\hbar\omega=1.55$~eV. The
symbol $\lambda$ is the wavelength here and in the tables, never $k\cdot u$.
Since
$\kappa$ is conserved in a plane wave, $\chi$ is fixed by the initial energy
and does not drift during the pulse. In the focused geometry $\kappa$ is not
conserved, and the estimate is low by roughly $\Delta u_{\perp}/\gamma$.

The three limits on the classical treatment are the classical radiated
fraction $R_{C}\simeq\tfrac23\alpha a_{0}\chi N$, the number of emitted
photons $N_{\gamma}\sim\alpha a_{0}N$, and the radiative spin-flip probability
$P_{\rm sf}\sim\alpha a_{0}\chi^{2}N$ at $\chi\ll1$. The first two are from
Ref.~\cite{DiPiazza2012} and the third from
Refs.~\cite{Seipt2018,Li2019}, with the reviews
\cite{Gonoskov2022,Fedotov2023} for context. The classical description
dominates while $\chi\ll1$, $R_{C}\ll1$, and $P_{\rm sf}$ stays below the
geometric signal $\sin^{2}(\Theta_{\rm net}/2)$.

Table~\ref{tab:Sregime} evaluates these on the plane-wave grid. Since
$\Theta_{\rm net}\propto a_{0}^{2}N$ and does not depend on $\gamma$ at all,
while all three limits tighten with $\gamma$, the optimum runs against
intuition, and the practical form is this: keep the electron slow, pay in
intensity and duration, and accept the best classical point on the grid, which
is circular polarization at $\gamma\simeq1$--$10$, $a_{0}\simeq50$--$75$ and
$N\simeq16$--$32$, where $\Theta_{\rm net}$ reaches $0.1$--$0.4$~rad with
$\chi\simeq2\times10^{-4}$--$5\times10^{-3}$ and
$R_{C}\lesssim1.3\times10^{-2}$.

\begin{table*}[!tbp]
  \caption{Applicability of the classical treatment on the plane-wave grid,
    circular polarization, Gaussian envelope, $800\,\mathrm{nm}$.
    $\Theta_{\rm net}$ is measured; $\chi$, $R_{C}$, $N_{\gamma}$ and
    $P_{\rm sf}$ are estimated from the scalings cited in the text and were
    not integrated.}
  \label{tab:Sregime}
  \begin{ruledtabular}
  \begin{tabular}{cccccccl}
    $\gamma$ & $a_{0}$ & $N$ & $\Theta_{\rm net}$ (rad) & $\chi$ & $R_{C}$
      & $P_{\rm sf}$ & Verdict \\
    \hline
    $1$ & $10$ & $8$ & $1.8\times10^{-3}$ & $3.0\times10^{-5}$
      & $1.2\times10^{-5}$ & $5\times10^{-10}$
      & classical; signal negligible \\
    $10^{3}$ & $10$ & $8$ & $1.8\times10^{-3}$ & $6.1\times10^{-2}$
      & $2.4\times10^{-2}$ & $2\times10^{-3}$
      & radiative flip exceeds the geometric one \\
    $5\times10^{3}$ & $10$ & $8$ & $1.8\times10^{-3}$ & $0.30$
      & $0.12$ & $5\times10^{-2}$ & quantum; BMT inapplicable \\
    $1$ & $75$ & $32$ & $4.0\times10^{-1}$ & $2.3\times10^{-4}$
      & $2.7\times10^{-3}$ & $9\times10^{-7}$
      & best classical point of the grid \\
    $10$ & $75$ & $8$ & $1.0\times10^{-1}$ & $4.5\times10^{-3}$
      & $1.3\times10^{-2}$ & $9\times10^{-5}$
      & classical; more realistic beam \\
    $10^{2}$ & $75$ & $8$ & $1.0\times10^{-1}$ & $4.5\times10^{-2}$
      & $0.13$ & $9\times10^{-3}$
      & radiated fraction $13\%$; boundary \\
    $10^{3}$ & $210$ & $10$ & $0.99$ & $1.27$ & $13$ & $25$
      & $\Theta\sim1$ only deep in the quantum regime \\
  \end{tabular}
  \end{ruledtabular}
\end{table*}

The working point is not free of qualifications, and none of them is
numerical. At $a_{0}=75$ and $N=32$ the estimated photon number is
$N_{\gamma}\simeq17$, so the spin state is not strictly pure. Each photon
carries $\lesssim10^{-3}$ of the energy and the decoherence is small, but the
purity assumption has to be stated. An intensity of
$1.2\times10^{22}\,\mathrm{W\,cm^{-2}}$ at $800\,\mathrm{nm}$ with $32$ cycles
is also a demanding pulse, and whether the required synchronization is
realistic is an experimental question we have not addressed.

The third caveat is the one whose data we already have. Every number in
Secs.~\ref{sec:S4} and~\ref{sec:S5} is for a single electron at a fixed impact
parameter, and no average over the transverse profile of the beam, over the
focal volume or over the spectrum was performed. That is not a small omission
in a focused geometry, because the measured spread of $|r_{0}|$ over impact
parameters within one waist is a factor of $20$, and the CEP modulation of
$|r_{0}|$ off axis reaches $35\%$ at $b=1.5w_{0}$ for linear polarization
(Sec.~\ref{sec:S5.6}). A realized measurement therefore samples a distribution
of rotations, and converting the numbers below into
an expected polarization requires that average, which we did not compute.

The focused geometry is more forgiving on radiation and less forgiving on
everything else. At $a_{0}=24$, $\gamma=10$, $N=1$ and $\varepsilon=0.15$ the
same scalings give $\chi=1.5\times10^{-3}$, $R_{C}=1.7\times10^{-4}$,
$N_{\gamma}=0.18$ and $P_{\rm sf}=3.7\times10^{-7}$ (estimated; that
combination is not among the deposited runs), so radiation reaction changes the
result by $\lesssim0.02\%$. The binding constraint there is ponderomotive
ejection instead: $\Delta u_{\perp}=0.52$ at that point, a deflection of
$0.05$~rad, which is tolerable, while $\Delta u_{\perp}=4.0$ at $a_{0}=8$ and
$\gamma=1$, which is not.

\section{What breaks the theorem}
\label{sec:S7}

Table~\ref{tab:Sbreak} separates the mechanisms that leave
Eq.~(\ref{eq:Sresult}) intact from those that destroy it. The first group was
tested directly. The second was tested only for focusing and for a pulse with
nonzero electric area. Radiation reaction, photon emission, a standing wave, a
medium and a static field were not integrated, and their entries are estimates
built from the scalings of Refs.~\cite{DiPiazza2012,Gonoskov2022}.

The pattern behind the table is simple. Anything that preserves both
$A=A(\eta)$ and $u(+\infty)=u_{0}$ leaves the holonomy alone, because those
two facts are the entire input to Secs.~\ref{sec:S2.1} and~\ref{sec:S2.2}.
Anything that supplies a second null direction, gives $k$ a mass, or makes the
electron lose energy breaks one of them.

Of the entries in the table, radiation reaction deserves the longest remark,
because it does more than shift a number. The Landau--Lifshitz force carries
the classical electron time $\tau_{0}=2e^{2}/3mc^{3}$, and once a dimensionful
constant sits in the equation of motion the argument of Sec.~\ref{sec:S3.2}
collapses, since $\tau_{0}\omega$ and $\gamma$ can then be formed and the
answer is free to depend on both. Its weight can be estimated from the radiated
fraction, a relative distortion $2R_{C}$ of the curve traced in the rest frame
against a holonomy of relative size $\ano a_{0}$, so that
$\Theta_{\rm RR}/\Theta_{\rm net}\sim2R_{C}/(\ano a_{0})$. Evaluated on the
rows of
Table~\ref{tab:Sregime} that estimator is $2\times10^{-3}$ at $\gamma=1$,
$a_{0}=10$, $N=8$ and about $4$ at $\gamma=10^{3}$, where the correction has
overtaken the quantity it corrects. It is also $6\times10^{-2}$ at the best
classical point of that table and $0.30$ one row below it, at $\gamma=10$,
$a_{0}=75$, $N=8$. A correction of six to thirty percent sits badly with a law
quoted to a percent, and only an integration of the Landau--Lifshitz equation
would settle the matter. We did not perform one.

The published mechanism closest to this one is the rotation of the spin toward
the propagation axis by the anomalous moment in a circularly polarized
field~\cite{Li2022}. That effect is linear in the anomaly, and the reason is
instructive here: repeated emission re-prepares the transverse spin with a
phase locked to the field, so the cancellation of Sec.~\ref{sec:S2.2} never
gets the closed loop it needs. Switch the emission off and the linear term
disappears again.

The first order also returns with no radiation involved at all. A
counterpropagating wave introduces a second null direction $k_{2}$, and the two
little algebras $\mathfrak{iso}(2)_{k_{1}}$ and $\mathfrak{iso}(2)_{k_{2}}$
together generate the whole of $\mathfrak{so}(1,3)$, so the ordering in
Eq.~(\ref{eq:Sgenerator}) survives even at
$g=2$~\cite{Bauke2014a,Rosanov2021}. A medium does it more cheaply. The
dispersion relation gives $k$ a mass, $k^{2}=\omega_{p}^{2}$, the little group
stops being $ISO(2)$, and the nilpotency on which Eq.~(\ref{eq:Slambda0}) rests
is gone. This is the effect Tikhomirov identified~\cite{Tikhomirov2003}, and
its weight relative to the holonomy is $(n_{e}/n_{c})/(\ano a_{0})$, already of
order $0.5$ at $n_{e}/n_{c}=6\times10^{-3}$ and $a_{0}=10$. Residual gas at
that density competes with the effect on equal terms.

A dressed anomaly $\ano(\chi)$ leaves the structure intact, and it turns the
result into a measurement. Since $\Theta_{\rm net}$ is quadratic in $\ano$ and
no
other plane-wave mechanism produces a net rotation, the transverse
polarization angle measures the square of the field-dressed anomalous
moment~\cite{Ilderton2020,Torgrimsson2021} with no $g$-independent background
to subtract. One qualification is internal to the derivation. The cancellation
of Sec.~\ref{sec:S2.2} takes $\ano$ outside the integral, so a phase-dependent
$\ano(\chi(\eta))$ would leave $\int \ano(\eta)\,\nhat\wedge
d\bm{a}_{\perp}\neq0$
and restore a first-order term. At the working point $\chi\simeq2\times10^{-4}$
the dressing is far too weak for that to matter, but the statement is exact
only for constant $\ano$. How the area law is modified when $\chi$ approaches
unity, where $\ano(\chi)$ is not defined outside the locally constant field
approximation, we did not analyze.

\begin{table*}[!tbp]
  \caption{Mechanisms that do and do not break Eq.~(\ref{eq:Sresult}). Rows
    marked measured were computed here; rows marked estimated rest on
    published scalings and were not integrated. $\Theta_{\rm pw}$ denotes the
    plane-wave holonomy $-\tfrac12\ano^{2}\Anet$.}
  \label{tab:Sbreak}
  \squeezetable
  \footnotesize
  \begin{ruledtabular}
  \begin{tabular}{llcl}
    Mechanism & Breaks it? & Order in $\ano$ & Size \\
    \hline
    Transverse initial momentum, up to $u_{0\perp}=17$
      & no & $\ano^{2}$
      & angle identical to 9 digits, only the axis moves (measured) \\
    Copropagating geometry, $\kappa=0.025$
      & no & $\ano^{2}$ & same angle (measured) \\
    $\int \bm{a}_{\perp}d\eta\neq0$
      & no & $\ano^{2}$
      & law holds to $10^{-5}$ relative at $\Anet=-0.14159$ (measured) \\
    Reparameterization of the phase at fixed curve
      & no & $\ano^{2}$
      & 8--9 matching digits (measured) \\
    Chirp and envelope shape
      & no & $\ano^{2}$
      & enter through $\Anet$ alone; see Sec.~\ref{sec:S3.1} (measured) \\
    Rotating polarization plane
      & no & $\ano^{2}$
      & law holds through $\Anet=1.5459$ to 6 digits (measured) \\
    Carrier-envelope phase
      & no & $\ano^{2}$
      & exact for circular, $\le2.4\times10^{-10}$~rad elliptical
        (measured) \\
    Initial spin orientation
      & no & $\ano^{2}$
      & net operator unchanged, projection changes (measured) \\
    \hline
    $\bm{a}_{\perp}(+\infty)\neq \bm{a}_{\perp}(-\infty)$
      & yes & $\ano^{0}$
      & $\|R-\mathbb{1}\|=3.46$ already at $g=2$ (measured) \\
    Finite focusing, $\varepsilon=1/kw_{0}$
      & yes & $\ano^{0}$
      & all coefficients $\propto\varepsilon^{2}$, exceeds $\Theta_{\rm pw}$
        at $w_{0}\lesssim5\lambda$ (measured) \\
    Radiation reaction
      & yes & $\ano^{1}$
      & $2R_{C}/(\ano a_{0})$: $2\times10^{-3}$ at $\gamma=1$, $\sim4$ at
        $\gamma=10^{3}$ (estimated) \\
    Photon emission
      & yes & $\ano^{1}$
      & scales with $N_{\gamma}\sim\alpha a_{0}N$ (estimated) \\
    Standing wave
      & yes & $\ano^{0}$
      & gain of order $1/\ano^{2}\sim10^{6}$, orbit chaotic (estimated) \\
    Plasma, $v_{\rm ph}\neq c$
      & yes & $\ano^{1}$
      & $(n_{e}/n_{c})/(\ano a_{0})\sim0.5$ at $n_{e}/n_{c}=6\times10^{-3}$
        (estimated) \\
    Static axial field on top of the wave
      & yes & $\ano^{1}$
      & the one case where $\ano\gamma$ returns,
        $\Theta\sim \ano\gamma\,\omega_{c}t$ (estimated) \\
  \end{tabular}
  \end{ruledtabular}
\end{table*}

%% ---------------------------------------------------------------------------
%%  BIBLIOGRAPHY, written out.  Produced by BibTeX from manuscript/refs.bib
%%  with apsrev4-2.bst and pasted in, so that this file builds with pdflatex
%%  alone.  To change a reference, edit manuscript/refs.bib and regenerate.

\end{document}